\documentclass{article}
\usepackage[utf8]{inputenc}
\usepackage{mathtools}
\usepackage{latexsym,float}
\usepackage{authblk}
\usepackage{epsfig}
\usepackage{epstopdf}
\usepackage{amsmath,hyperref}
\usepackage{amssymb}
\usepackage{graphicx}
\usepackage{eepic}
\usepackage{bbm}
\usepackage{textcomp}
\title{Cosmological bounce and the cosmological constant problem}

\begin{document}
\author[1,2]{Petar Pavlović \thanks{petar.pavlovic@kozmologija.org}}
\author[1,3]{Marko Sossich \thanks{marko.sossich@fer.hr}}
\affil[1]{\textit{\small{Institute for Cosmology and Philosophy of Nature,
Trg svetog Florijana 16, Križevci, Croatia}}}
\affil[2]{\textit{\small{Department of Physics,
Ramashna Mission Vivekananda Educational and Research Institute,
Belur Math 711202, West Bengal, India}}}
\affil[3]{\textit{\small{University of Zagreb, Faculty of Electrical Engineering and Computing, Department of Physics,
	Unska 3, 10 000 Zagreb, Croatia}}}

\maketitle
\begin{abstract}
We discuss how the modifications of the standard Einstein's equations needed to support the cosmological bounce can at the same time lead to vanishing of the well known cosmological constant problem, while also studying the effects of spacetime fluctuations. We first concentrate on the case of the classical FRWL spacetime and show that in a bouncing cosmology, where the modification of the Einstein-Hilbert action can be represented by the most general function needed to support the bounce, the cosmological constant problem is absent if this function at late cosmological times approaches a sufficiently large value. We show that this result is general and does not depend on the details of modifications needed to support the cosmological bounce. Therefore, the bouncing cosmologies could potentially at the same time solve the problem of singularity and cosmological constant. Motivated by the recent studies of cosmological constant problem in the context of fluctuations of the metric, we then generalize our study to incorporate the effects of spacetime fluctuations. We show  that some problems of the recent earlier proposals, like singularities and negative values of the scale factor, are also naturally resolved in the approach proposed here. 
\end{abstract}
\section{Introduction}
The standard $\Lambda$CDM model of cosmology, based on general theory of relativity and assumptions of large scale homogeneity and isotropicity, was tested and shown to be consistent with numerous observations. Some of these measurement include growth of cosmological observations, 
supernova and microwave background measurements, abundances of chemical elements and consistency of of the age of the Universe with the age of astronomical objects \cite{tests}. However, this model is at the same time plagued with various fundamental problems. The problem of the existence of singular solutions can actually not be resolved within the assumptions of the standard cosmological model, since the existence of singular solutions is actually a necessary feature of general relativity (as long as the usual conditions on matter-energy are assumed), as proven by the theorems of Hawking and Ellis \cite{hawk1, hawk2, hawk3}. The existence of initial singularity signifies that the physical theory is incomplete and therefore should not be trusted in the regimes of the very early Universe. For this reason it is not justified to treat the initial "big-bang" singularity as a physical reality -- which would mean that physics needs to abandon the goal of a complete and consistent description of the origin of the Universe, but it should rather be viewed as the mathematical pathology signalling the limitations of general relativity. In our view, all other open problems of the standard cosmological model should be viewed from the same perspective. \\ \\
The most serious of further problems include the dark matter problem \cite{dm1,
dm2}, the horizon problem \cite{linde}
 and the dark energy problem -- which under the assumptions of the 
$\Lambda$CDM model and the Standard model of particle physics leads to the cosmological constant fine-tuning problem
\cite{k1, k2, k3, k4}. Since we still do not have a proper candidate for a theory of quantum gravity, which would hopefully solve these problems, we believe that a reasonable approach consists in considering a broad class of natural extensions of general relativity which are capable of resolving such issues with the minimal number of additional assumptions. We will present a proposal for such a set of assumptions in section II. In this work we will concentrate only on the problem of the cosmological constant from such an effective approach towards quantum gravity. However, numerous contributions to other mentioned problems from the similar perspective can be found in the literature, and we will refer to some of them in section II. 
\\ \\
The special importance of the cosmological constant problem comes from the fact that it arises when the fundamental results of general relativity and quantum mechanics are contrasted to each other. Namely, the quantum mechanics predicts a huge value of the vacuum energy density, which would then lead to a huge value of the constant cosmological term determining the evolution of the Universe. But what is empirically found is that the evolution of the Universe could be consistent with the general relativity only if such term is 122 orders of magnitude smaller that the above mentioned value predicted by the quantum mechanics.
\\ \\
In our view, this problem should not be considered separately, but together with other fundamental problems of the standard cosmological model -- the singularity and horizon problem, and the solution to all these problems should be searched for in one unified and consistent theoretical framework. We will try to sketch the basic features of such a framework in Section II. The most important feature of this framework is that it requires the extension of the gravitational field equations with the additional correction terms. In this framework we approach the cosmological constant problem in the following manner: the observational value of the cosmological term is the result of the two following contributions: the quantum-mechanical vacuum energy density and the correction terms to the standard gravitational Lagrangian, which of course change their values during the cosmological evolution. Therefore the effective "cosmological term" consisting of these two contributions is no longer a constant, and under suitable conditions it reaches the required small value characterizing the Universe today. In this respect, the approach discussed here shares the same logic with various proposals for resolution of the cosmological constant problem using modified gravity and dynamic dark energy \cite{stefancic,sola, sola2}. However, the basic differences, with respect to the earlier proposals are : a) The question of the interplay between the vacuum energy density and modification of equations for gravity is discussed in the setting of a non-singular bouncing cosmology, which determines the qualitative features of such corrections, b) The treatment is kept general and not limited by the assumptions of any specific model of modified gravity. In this framework we obtain a general and simple result: if the modification of the Lagrangian supports the bouncing cosmology and the corresponding terms reach some sufficiently large value at late cosmological times, then the corresponding cosmological solutions will include the phase in which the measured effective cosmological term reaches arbitrary small positive values.\\ \\
In the second part of our work we generalize our treatment to include the effect of spacetime fluctuations. It was very properly pointed in \cite{carlip} and \cite{unruch}, as well as in some earlier investigations of this topic in \cite{unruch2} and \cite{unruch3}, that at small (Planck) scales,  one should not take the usual, highly idealized FRWL metric as a justified 
approximation. In other words, due to the fact that at such scales the spacetime should be expected to wildly fluctuate, it is problematic to analyse the question of cosmological constant in the framework of the large scale FRWL metric, which is assuming homogeneity and isotropicity. It was argued in \cite{carlip} and \cite{unruch} that this fact, implying the usage of a more general metric - containing the quantum fluctuations of both vacuum energy density and spacetime -- can explain the cosmological constant problem, since the large value of the cosmological constant can become hidden in the quantum fluctuations (however, note that the proposals made in \cite{carlip} and \cite{unruch} for the realization of this idea are quite different in their nature). However, we are here not concerned with the details and potential difficulties of the idea of fluctuating vacuum density, since we will study only the consequences of the fluctuations of the spacetime metric. The idea of fluctuating spacetime metric was actually originally proposed by Wheeler in his concept of the spacetime foam \cite{wheeler}.  \\ \\
Since we have, in the first part of the paper, already discussed how the cosmological constant problem can in principle be solved in the very general type of modified gravity bouncing cosmological model -- based on the classical FRWL metric -- we are not primarily interested in searching for the resolution of the cosmological constant problem using the quantum fluctuations. We are, first of all, interested in generalizing our discussion to incorporate the effects of spacetime fluctuations and to analyse its consequences. With this aim, we study the  spacetime fluctuations while ignoring the vacuum fluctuations. We start from the very general metric and study the problem within the context of our assumptions stated in Section II. Our results will, however, show that the idea of hiding the huge value of cosmological constant in the spacetime fluctuations can be further supported and some of the important difficulties of earlier proposals can be fixed in the framework we will develop in this work. 
\\ \\
We first show that, under a certain condition for the values of the shear function, the effect of spacetime fluctuations will not lead to oscillations around the cosmological bounce. Indeed, if the effects of spacetime fluctuations are small enough, they will just lead to a small correction to the classical bouncing solutions. We next show that in the context of our model, where the cosmological term is no longer a constant, there exists a much richer set of cosmological solutions. The main difference with respect to the cosmological constant case, discussed in \cite{unruch}, is that there is now also a possibility of singularity free oscillations, including the oscillations with the growing amplitude. Such solutions have all of the desired properties discussed in \cite{unruch} for hiding the large cosmological constant, but at the same time they do not suffer from the pathological singularities at $a=0$ and always have the positive value of the scale factor. In addition, we have also considered some other types of singularity free solutions, including the loop quantum cosmology motivated solutions, and provided the analytical reconstruction of singularity free solutions. \\ \\
This work is organized as follows: in section II we state the basic assumptions of our approach, in section III we consider the classical FRWL spacetime 
  and discuss the absence of the fine-tuning problem in bouncing cosmologies, in section IV we consider the quantum fluctuations of spacetime, in section V we discuss the cosmological constant limit of the spacetime fluctuations, in section  VI we investigate the absence of oscillatory solutions around the cosmological bounce, we present different classes of singularity free cosmological solutions in section VIII and finally discuss our findings in section IX.

\section{The assumptions of our approach }
We believe that the open problems in our understanding of the Universe and its evolution should be addressed in the framework which is based on the following set of assumptions, based on numerous works exploring the alternative and modified gravity models in the past decades :
\begin{enumerate}
    \item The description of cosmological evolution should be free of pathological features such as spacetime singularities. In fact, the removal of singularities contained in general relativity was for a long time one of the reasons for exploring alternative and modified theories of gravity in various settings
    \cite{hidekazu,ratbay, bojowald, nojiri, mudrac}
    \item For the reason that it necessarily predicts the existence of initial singularity in cosmology and that it ignores the uncertainty principle of quantum mechanics, the standard Einstein's general theory of relativity should be regarded as incomplete and inadequate for description of high-curvature regimes reached in the very early Universe. 
    \item While it is taken as an usual assumption, based on a simple dimensional analysis and not on strong theoretical reasons, to expect that gravity should be fully described in terms of a yet unknown quantum field theory above the Planck energy scale, there is no reason which would prevent the quantum corrections to the classical equations for gravity to become active even on a scales much smaller than the Plank scale. In fact, it is currently completely unclear if the Universe in its evolution ever reaches the Planck scale regime. In our understanding this question is not of central importance for this discussion at this point. Whatever be the highest energy scale reached in the evolution of the very early Universe, the demand for the absence of singularities requires that the corrections to the classical Einstein's equations are strong enough in this regime to remove the initial singularity. 
    \item The corrected form of Einstein's equation should be in accord with all currently confirmed results of $\Lambda$CDM model. 
    \item Taking into account the vast observational support for the expanding evolution of the Universe, which evolved through a radiation dominated, matter dominated and dark-energy dominated phases, and combining this picture with the stated assumptions 1.-- 4. naturally leads to the replacement of the big bang singularity with the cosmological bounce - a transition from the earlier phase of contraction into the phase of the expansion of the Universe, enabled by the corrections to the standard Einstein's equations and later followed by the usual phases of the standard cosmology. There is an extensive research literature on various types of bouncing solutions, constructed in different models and using various assumptions \cite{cai,roshan,salehi, pan, sahoo, bari, bajardi}. Recently, a model-independent and general approach for studying the bouncing and cyclic solutions was proposed in \cite{mi1} and \cite{mi2}. 
    \item In the situation where the basic cosmological questions are related to the regime of the very early Universe for which we should strongly suspect the validity of standard general relativity and for which we do not have the proper theory of quantum gravity, the logical approach is to first address these questions in a framework where all the basic physical assumptions of the standard Einstein's general relativity are preserved and only the mathematical form of the field equations is changed. This assumption agrees with the approach of various classes of modified gravity theory \cite{clifton}. Such type of generalizations can, for instance, already be motivated by considering the quantum effects -- such as the higher loop-corrections coming from the self-interaction of gravitons, which then introduce the higher order curvature invariants to the standard Lagrangian density for gravity. 
    \item From the methodological point of view, it is in general often possible to obtain some desired features of the theory by adding new \textit{ad hoc} elements, designed in a specific way to lead to the desired features. The problem is that, in this approach, physics becomes the game of adding newer and newer unmotivated ingredients and thus increasing the parameter space in order to obtain better fit of the theory to the data. As a consequence, a physical theory looses its power to describe the reality in a logical, motivated and coherent way -- which actually represents one of the main strengths of physics as a science. This is, apart from the existence of singularities, one of the main shortcomings of the $\Lambda$CDM model. For this reason we feel that the approach consisting in introduction of new exotic forms of fluids, scalar fields with specific forms of potential, new types of interactions or dimensions etc. -- not motivated by some independent deep theoretical reasons or independent empirical justifications, but introduced just for the sake of obtaining a specific type of solution, is in its essence artificial and does not represent a proper methodology of research. 
\end{enumerate}
We will try to demonstrate that in the framework of these assumptions the central problems of the standard cosmology, like the cosmological constant problem, are simply resolved without the need to invoke any additional mechanisms. Similarly, we have also demonstrated in an earlier paper that the problem of creation of magnetic fields can also be naturally solved in this paradigm \cite{natty}.
The usual approach followed in investigation of modified gravity in the cosmological context is to choose a specific form of a modification of the standard Einstein's theory and then to solve the corresponding equations, obtain the cosmological solutions and discuss their properties. By definition, this approach has a limitation that the derived results are characteristic only for the concrete assumed model and not valid in general. As it is currently not known which of the possible numerous modifications are preferred by Nature, it is not clear what is the proper significance of such results. Motivated by such problems, we will approach the problem of the cosmological constant from a complementary perspective. We start from the presented assumptions 1.-- 7. and from them we reconstruct the features that the corrections to the standard Einstein's equations need to take in the cosmological setting. From this we obtain the approximate form of the corrections entering into the modified Friedmann's equations and discuss the issue of the cosmological constant problem. We first present the discussion on such type of reconstruction assuming the classical properties of spacetime in the following section, and then after that, we consider the problem from the perspective of quantum induced spacetime fluctuations. 
\section{Classical FRWL spacetime}
Following the stated assumptions 2., 3., 6. and 7. we conclude that the appropriate form of the gravitational field action for the description of the high curvature regime reached in the early Universe can be represented by the following effective action:
\begin{equation}
    \mathcal{S}_{eff}= \frac{1}{16 \pi G}\int\sqrt{-g} Rd^{4}x + \mathcal{S}_{mod},
    \label{akcija}
\end{equation}
Where $\mathcal{S}_{mod}$ is a general type of correction to the classical action of gravitational field which supports the bouncing cosmological solutions. It can in principle be some arbitrary function involving invariant quantities such as Ricci scalars and contractions of Ricci and Riemann tensors and their derivatives. Different types of such modifications are investigated in various modified gravity theories and they can also arise due to quantum corrections (for instance, due to the self-coupling of graviton). After adding the action for matter fields and varying with respect to metric, one obtains the effective field equation for gravity:
\begin{equation}
    G_{\mu \nu}^{eff}= G_{\mu \nu} + G_{\mu \nu}^{mod}= 8 \pi G
    (T_{\mu \nu}^{rad, mat}+ T_{\mu \nu}^{vac}),
    \label{mod}
\end{equation}
here $G_{\mu \nu}$ is Einstein's tensor, $G_{\mu \nu}^{mod}$ simply denotes the collection of terms which arise after variation of $\mathcal{S}_{mod}$ with respect to the metric, $T_{\mu \nu}^{rad, mat}$ is stress-energy tensor for matter and radiation and $T_{\mu \nu}^{vac}$, is the stress-energy contribution of the quantum vacuum. We keep the discussion completely general and do not further specify $G_{\mu \nu}^{mod}$, apart from the demands: i) that $G_{\mu \nu}^{mod}$ represents a proper tensor, and ii) that the complete stress-energy tensor stays conserved. The second requirement implies $\nabla^{\mu}G_{\mu \nu}^{eff}=\nabla^{\mu}G_{\mu \nu}^{mod}=0$. 
\\ \\
Despite the mentioned general nature of $G_{\mu \nu}$ it is simple, as well as essential for the further discussion, to note that if the spacetime is assumed to be homogenous and isotropic then the only independent cosmological variable on which this collection of terms can depend is time. Therefore, on such a spacetime the components of $G_{\mu \nu}$ need to be equivalent to some functions of time. In accord with this, we assume the classical FRWL spacetime can still be used as a justified approximation for description of cosmological evolution, and thus take the metric to be given by
\begin{equation}
 ds^{2}=-dt^{2}+a(t)^{2}\bigg( dr^{2}+ r^{2}(d\theta^{2} +  \sin^{2} \theta d\phi^{2})  \bigg).
 \label{flrwmetrika}
\end{equation}
We now obtain the modified Friedmann's equations resulting from equations (\ref{mod})--(\ref{flrwmetrika}):
\begin{equation}
3H^{2}= 8 \pi G \bigg(\rho_{rad}^{0} a^{-4}+ \rho_{mat}^{0}a^{-3} + \rho_{vac} \bigg) + S(t) .
\label{modfri1}
\end{equation}
\begin{equation}
    \frac{\ddot{a}}{a}= - \frac{4 \pi G }{3}
    \bigg(\rho_{mat} + \rho_{rad} + 3p_{rad} + \rho_{vac}+3p_{vac} \bigg) + \frac{G_{rr}^{mod}(FRWL)}{2a^{2}}- \frac{S(t)}{6}
    \label{modfri2}
\end{equation}
Here we have introduced $S(t)=-G_{00}^{mod}(\textit{FRWL})$, $T_{00}^{vac}=-\rho_{vac}g_{00}$ and $T_{ij}^{vac}=-p_{vac}g_{ij}$, while also taking the standard assumption of matter and radiation modelled by the ideal fluid with the equation of state $w=0$ and $w=1/3$ respectively. Demanding that $\nabla^{\mu}G_{\mu \nu}^{mod}=0$ we obtain the identity for 
$G_{r r}^{mod}$ on FRWL spacetime:
\begin{equation}
\frac{3H}{a^{2}}G_{r r}^{mod}(FRWL)=\dot{S}(t)+ 3H S(t),
\end{equation}
while all other non-vanishing components of $G_{i j}^{mod}$ here need to be simply related to $G_{r r}^{mod}$ by multiplication of the corresponding metric coefficients, by the virtue of spherical symmetry of the considered spacetime. Since the equation (\ref{modfri2}) is, under these assumptions, just a time derivative of (\ref{modfri1}), as can be easily checked, we will in further analysis inspect mostly equation (\ref{modfri1}). \\ \\
The reader should understand that modified Friedmann's equations (\ref{modfri1}) and (\ref{modfri2}) do not come as some arbitrary introduction of function $S(t)$ by hand at the level of the field equations for gravity, but they follow as the most general form of field equations from the effective action (\ref{akcija}) on the FRWL spacetime. 
We note that, despite their simplicity, the presented equations are indeed very general, as different classes of possible gravity theories -- such as Einstein's gravity, $f(R)$ gravity, $f(T)$ gravity, Loop quantum cosmology and various modifications of the stress-energy tensor etc. - are simply obtained by the proper choice of the $S(t)$ function. The clear advantage of this approach is that a very broad class of possible modifications is represented by a single function which measures a departure from the Einstein's equations in the case of this spacetime geometry -- and which is thus very appropriate for confrontation with  experiments. To put it in another words, the advantage of this perspective is that one does not need to be limited to any of the numerous possible modifications of Einstein's gravity, but can approach the physics beyond standard general relativity in a model-independent manner. Then the results of this approach can in principle, if needed, be converted in the language of some specific theory of gravity by reconstruction. On the other hand, the obvious disadvantage of this approach is that the conclusions obtained for $S(t)$ are valid only for the FRWL spacetime geometry. However, it seems reasonable to expect that a similar investigation of $G_{\mu \nu}^{mod}$ for various spacetime geometries of physical interest -- focused on the overcoming of limitations of standard general relativity in agreement with experiments, could provide a better understanding of the paths towards quantum gravity. 
\\ \\
In the standard $\Lambda$CDM cosmology (which violates the assumptions 1. and 7. it follows $S(t)=\lambda= constant$. Since the contribution $\lambda_{eff}= \lambda + 8 \pi G \rho_{vac}$ can be determined from the observations, by measuring the values of all other parameters appearing in (\ref{modfri1}), and $\rho_{vac}$ can be estimated from the quantum field theory considerations, $\lambda$ needs to take the specific value to be consistent with these results. It then follows that this constant, although theoretically independent of $\rho_{vac}$, needs to be mysteriously very fine-tuned to lead to the empirical result for $\lambda_{eff.}$. Namely, if the standard effective field theory estimate is used, then it follows that the value of the vacuum energy can be expected to be $\rho_{vac} \sim 1$, in Planck units. On the other hand, by measuring the expansion of the Universe the observations seem to demand that $\lambda_{eff}=5.6 \cdot 10^{-122}$, in the same units. In this framework, it seems that the only possibility is that $\lambda$ needs to be  so finely chosen that it cancels the contribution of $\rho_{vac}$ to $122$ decimal places. It is completely unclear how this could be possible, since both constants have a completely different nature: $\rho_{vac}$ represents vacuum energy density, and $\lambda$ fundamentally corresponds to a constant term in the action integral for gravity added to the Ricci scalar.  Unlike the $\Lambda$CDM case, we will soon show that there is no such fine-tuning problem in the approach based on the presented assumptions 1.-- 7., i.e. when $\lambda_{eff}=S(t)+ 8 \pi G \rho_{vac}$, with $S(t)$ being an appropriate, non-constant function. 
\\ \\
We now turn to the inspection of different cosmological regimes in order to reconstruct the proper time dependence of $S(t)$ in them, in the light of our starting assumptions. After obtaining the necessary constraints on the function $S(t)$ in order to obtain the cosmological bounce, we study the consequences of those conclusions on the cosmological constant problem.

\subsection{Cosmological bounce}

Taking the assumption 5., we consider the replacement of the initial cosmological singularity with the bounce, i.e. the transition from the previous contracting phase into the current expanding phase. At the bounce, $t=t_{b}$, where the scale factor reaches its minimal value, $a=a_{min}$, we have $H=0$ and therefore using (\ref{modfri1}) we obtain the following condition: 
\begin{equation}
S(t_{b})=- 8\pi G \bigg(\rho_{rad}^{0} a_{min}^{-4}+ \rho_{mat}^{0}a_{min}^{-3} + \rho_{vac} \bigg) .   
\label{sbounce}
\end{equation}
In this work we assume the validity of the standard estimate for the expectation value of the quantum vacuum, $\rho_{vac} \sim 1$ (in Plack units). Therefore, it necessarily follows that $S(t_{b})<0$ and $\rho_{vac} < 
 |S(t_{b})|$. Furthermore, since $t=t_{b}$ represents the minimum of the scale factor, around the bounce 
$a(t) \approx a_{min} + (C/2)(t-t_{b})^{2}$, where $C$ is a positive constant. From this, one easily obtains the approximation for $H(t)$ around the bounce. Therefore, the approximate form of $S(t)$ around the bounce is given by
\begin{equation}
\begin{split}
S(t)_{bounce} \approx 3 \bigg(\frac{C(t-t_{b})}{a_{min}+C (t-t_b)^{2}/2} \bigg)^{2} - 8 \pi G \Bigg[
\rho_{rad}^{0} \bigg(a_{min} + C(t-t_{b})^{2}/2 \bigg)^{-4} 
\\ \\
+\rho_{mat}^{0} \bigg(a_{min} + C(t-t_{b})^{2}/2 \bigg)^{-3} + \rho_{vac} \Bigg ]
\label{saroundbo}
\end{split}
\end{equation}
\subsection{Late time acceleration phase, matter and radiation phase}
From the assumption 4. we conclude that the form of $S(t)$ needs to lead to the late accelerated expansion phase, where the evolution of the Universe can be approximately described in the context of standard Einstein's equation, with the addition of a small positive cosmological constant (thus in contradiction with the discussed value of $\rho_{vac}$). As the late time accelerated expansion asymptotically approaches De Sitter solution, we approximately have $H(t) \approx H_{late}=constant$.  In our approach, using equation (\ref{modfri1}), neglecting the contribution of matter and radiation which becomes insignificant in this phase, we have that in the late acceleration phase
\begin{equation}
 S(t)_{late} \approx 3 H^{2}_{late} -  8 \pi G \rho_{vac} .
 \label{sac}
\end{equation}
From observations we can deduce that clearly $S(t)_{late}<0$, while also $S({t_{b}}) \ll S(t)_{late}$. The second inequality is obvious if, for instance, we take into account that, assuming the bounce happens prior to or comparable with the scale usually associated with inflation, $a_{min}<10^{-32}$. In any case, even if this assumption is relaxed, the contributions of matter and energy in the very early Universe (thus around the bounce) need to become huge and the last inequality can safely be assumed to necessarily hold. 
\\ \\
Again following the assumption 4., we demand that before entering into the discussed late acceleration phase the Universe passes through the stages of radiation and matter domination. This will be achieved 
if there is an interval of time for which 
\begin{equation}
 S(t)_{rad,mat} \approx - 8 \pi G \rho_{vac}  .   
 \label{smatrad}
\end{equation}
Since $S(t)_{bounce}<S(t)_{rad,mat}<S(t)_{late}$, it is obvious that the Universe whose evolution is guided by equation (\ref{modfri1}) will necessarily pass through the era of radiation and matter domination as long as $S(t)$ is a continuous function and the Universe is evolving between the bouncing and the late accelerated phase. The duration of these phases and further details of the cosmological evolution will of course dependent on the concrete form of correction term $S(t)$. In the same spirit it is easy to see that the problem of cosmological constant is generically absent in this approach. 
\subsection{The absence of the fine-tuning problem}
Basing ourselves on the discussion in the previous sections we can state the following general claim:
\\ \\
\textbf{If the function $S(t)$, describing the corrections to the classical Einstein's equations on FRWL spacetime, defined as $S(t)=-G_{00}^{eff}(FRWL)$, where $G_{00}^{eff}$ represents the general collection of terms arising from  modifications of classical action integral for gravity, is assumed to have the following properties
\begin{itemize}
    \item[i)] is a continuous function of time,
    \item[ii)] leads to the cosmological bounce,
    \item[iii)] at some late time approaches some value $S_{asymp}$ such that $3H_{0}^{2} - 8 \pi G \rho_{vac}<S_{asymp}$ (with $H_{0}$ being the Hubble parameter today),
\end{itemize}
 then the corresponding cosmological solutions will include the phase in which the measured effective cosmological term, $  \lambda_{eff}=S(t) + 8 \pi G \rho_{vac}$, reaches arbitrary small positive values. Furthermore, such solutions will also have the phase of radiation and matter domination}. \\
\\
\textbf{Proof}: From the equations (\ref{sbounce}), (\ref{smatrad}) and the condition $3H_{0}^{2} - 8 \pi G \rho_{vac}<S_{asymp}$ it follows, as previously discussed, that $S(t)_{bounce}<S(t)_{rad,mat}<S_{asymp}$ . Then, since the function $S(t)$ is a continuous function which takes the values in the interval from the $S(t)_{bounce}$ to the $S_{asymp}$ it will also necessarily reach the value $S(t=t')=- 8 \pi G \rho_{vac}$ (as demanded by the well known intermediate value theorem of mathematical analysis). Then, again from the continuous nature of $S(t)$, it follows that on some interval around $t'$ the function corresponding to the effective cosmological term encountered in observations, $\lambda_{eff}= S(t)+ 8 \pi G \rho_{vac}$, will be positive and arbitrary small. Furthermore, if on some interval of time after matter-radiation domination $S(t)$ can be approximated by (\ref{sac}) it will also lead to the late time accelerated expansion. \\ \\
Therefore, what in the standard cosmological model appears as the mysterious fine-tuning of two unrelated constants, $\rho_{vac}$ and $\lambda$, is here just the simple and natural consequence of the fact that the correction function $S(t)$, absorbing in itself the contribution from correction effects on the FRWL cosmology, needs to take all values in the interval between the bouncing phase and some moment in the late expansion of the Universe, $S_{asymp}$. This, of course, resolves the problem on the basic and conceptual level, while on the technical level the proper length of the cosmological interval corresponding to the observed value of the effective cosmological term, $\lambda_{eff}$, as well as the proper duration of matter-radiation domination, puts observational constraints on the values of the free parameters of the model and the possible form of the correction function $S(t)$. 
We can thus see that the well known problem of cosmological constant, has a simple and very general solution, based on a set of very minimal and logical assumptions (1.-- 7.), without any need to invoke modifications of quantum sector, introduce \textit{ad hoc} constructed hypothetical mechanisms or use bad metaphysics (i.e. scholastics) in the form of the so called anthropic principle. We should stress that we can of course not know from the current observations what is the possible upper boundary for $S_{asymp}$, as we can not empirically determine the future evolution of the Universe. Therefore, the value of $S_{asymp}$ can in the late Universe approach zero or even large positive values. In a sense, the larger the reached upper boundary for $S(t)$ is, the more far away from its present value close to $- 8 \pi G \rho_{vac}$ it is, and the less special the whole evolution appears. A specially consistent and complete type of scenario is given by cyclic cosmology – where after the discussed phases and the subsequent growth of $S(t)$, the correction function starts to decrease leading to a new bounce and beginning of a new cycle. 

\section{Fluctuations of the spacetime geometry}
In this section, we take into account the possibility that FRWL metric is not a proper approximation for the geometry of the Universe at the quantum scales and early times. This is due to the fact that in such regimes the spacetime can be expected to be wildly fluctuating. We thus follow and further generalize, introducing the corrections to Einsten-Hilbert Lagrangian, the discussion in \cite{unruch, unruch2, unruch3}.
In this section we study only the influence of spacetime fluctuations, neglecting the vacuum fluctuations and fluctuations of stress-energy in general.
As stated in Sec. 2 following the assumptions 2., 3., 6. and 7. the appropriate modification of the standard Einstein-Hilbert action should take the following form
\begin{equation}
      G_{\mu \nu}^{eff}= G_{\mu \nu} + G_{\mu \nu}^{mod}= 8 \pi G
    (T_{\mu \nu}^{rad, mat}+ T_{\mu \nu}^{vac}),
\end{equation}
where again we assume the stress-energy tensor conservation $\nabla^{\mu}G_{\mu \nu}^{eff}=0$. One can adopt the "GR picture" where the quantum corrections to the Einstein field equation can be expressed as a modified stress-energy contribution
\begin{equation}
     G_{\mu \nu}=8 \pi G
    (T_{\mu \nu}^{rad, mat}+ T_{\mu \nu}^{vac} + T_{\mu \nu}^{mod})=8\pi G T_{\mu \nu}^{eff},
\end{equation}
where $T_{\mu \nu}^{mod}\equiv -G_{\mu \nu}^{mod}/8\pi G$. In this convention, with the analogy of the perfect fluid stress-energy tensor from (\ref{modfri1}) and (\ref{modfri2}), the modified terms take the form
\begin{equation}
    \rho_{mod}=\frac{S(t)}{8\pi G},
    \label{rhomod}
\end{equation}
\begin{equation}
    p_{mod}=-\frac{1}{3H}\frac{\dot{S}(t)}{8\pi G}- \frac{S(t)}{8\pi G},
\end{equation}
where the stress-energy conservation takes the usual form $\dot{\rho}_{mod}=-3H(\rho_{mod} + p_{mod})$. Thus, the effective conservation law can be written as
\begin{equation}
    \dot{\rho}_{eff}=-3H(\rho_{eff} + p_{eff}),
\end{equation}
where
\begin{equation}
    \rho_{eff}=\rho_{mat}+\rho_{rad}+\rho_{vac}+\rho_{mod},
     \label{effrho}
\end{equation}
\begin{equation}
     p_{eff}=p_{rad}+p_{vac}+p_{mod}.
\end{equation}
If we consider the quantum regime of the spacetime fluctuations we can expect that
the FLRW metric will not hold anymore as the spacetime is becoming wildly fluctuating. At this point we make a following  assumption: we ignore the effects of quantum fluctuations on the functional dependence of modification terms to Einstein's equation and thus take them to be described in the same way as on the classical spacetime, by a function $S(t)$. When the spacetime describing the Universe becomes fluctuating, we need to start with the most general metric tensor $g_{\mu \nu}$. In the cosmological setting it is useful to decompose the metric tensor as follows 
\begin{equation}
    h_{ab}=g_{ab}+n_an_b.
\end{equation}
If $\Sigma_t$ is a spacelike Cauchy surface parametrized by $t$ and $n^a$ is a unit normal vector field to $\Sigma_t$ then the resulting metric tensor can be written as
\begin{equation}
    ds^2=-N^2dt^2 + h_{ab}(dx^{a}+N^adt)(dx^b+N^b dt).
    \label{metrika}
\end{equation}
If $t^a$ is a vector field satisfying $t^a\nabla_a t=1 $ which represents the "flow of time" through the spacetime, then $N=-t^an_a$ which is called the lapse function, $N^a=h^{ab}t_b$ are the shift vectors, and $h_{ab}$ is the induced spatial metric which depends on both time and space. Latin indices go from $1-3$. This is known as the 3+1 decomposition of the spacetime with spatial slices $t=constant$. In this decomposition the extrinsic curvature $K_{ab}$ of hypersurface $\Sigma_t$ is given by
\begin{equation}
    K_{ab}=\frac{1}{2N}\Big( \dot{h}_{ab} - D_aN_b - D_bN_a \Big),
    \label{exkurva}
\end{equation}
where $\dot{h}_{ab}$ is the time derivative of the spatial metric and $D_a$ is the derivative operator on $\Sigma_t$ associated with $h_{ab}$. The Einstein field equations are given by the set of equations
\begin{multline}
    \dot{K}_{ab}=-N\Big[ R^{(3)}_{ab} + KK_{ab} - 2K_{ac}K_{b}^c \\
    - 4\pi G h_{ab} \rho_{eff} - 8 \pi G\Big( T_{ab}^{eff} - \frac{1}{2}h_{ab}\mbox{tr} T^{eff}\Big) \Big] \\
    + D_aD_bN + N^cD_cK_{ab} + K_{ac}D_{b}N^c + K_{cb}D_aN^c,
    \label{glavna1}
\end{multline}
\begin{equation}
    R^{(3)}-K_{ab}K^{ab} + K^2=16 \pi G \rho_{eff},
    \label{hamconstr}
\end{equation}
\begin{equation}
    D^aK_{ab}-D_{b}K=-8\pi G J^{eff}_b,
\end{equation}
where $R_{ab}^{(3)}$ is the 3-dimensional Ricci tensor of $\Sigma$, $R^{(3)}=h^{ab}R_{ab}^{(3)}$ is the 3-dimensional Ricci scalar on $\Sigma$, $K=h^{ab}K_{ab}$, $\rho_{eff}=T_{ab}^{eff}n^an^b$ is the effective energy density given by (\ref{effrho}), $J_b^{eff}=-h_{b}^cT_{ca}n^a$ is the effective energy flux and $\mbox{tr} T^{eff}=h^{ab}T^{eff}_{ab}$. Taking the trace of (\ref{glavna1}) and combining it with (\ref{hamconstr}) it reduces to
\begin{equation}
    \dot{K}=-N\Big(K_{ab}K^{ab} + 4 \pi G (\rho_{eff} + \mbox{tr} T^{eff})\Big) + D^aD_aN + N^aD_a K.
    \label{kurva1}
\end{equation}
By splitting the extrinsic curvature $K_{ab}$ into the traceless part (commonly called shear tensor) $\sigma_{ab}$ and the trace part $K$ in the following manner
\begin{equation}
    K_{ab}=\sigma_{ab} + \frac{1}{3}Kh_{ab},
    \label{kurva2}
\end{equation}
 the time evolution of the shear tensor can be written as
\begin{multline}
     \dot{\sigma}_{ab}=-N\Big[ \frac{1}{3}K \sigma_{ab} - 2\sigma_{ac}\sigma_{b}^c + R_{ab}^{(3)} - \frac{1}{3}R^{(3)}h_{ab}\\
     -8\pi G\Big( T_{ab}^{eff} - \frac{1}{3}h_{ab}\mbox{tr} T^{eff}\Big) \Big] +
     D_aD_bN - \frac{1}{3}h_{ab}D^cD_cN \\
     +N^cD_c\sigma_{ab} + \sigma_{ac}D_bN^c \sigma_{cb}D_aN^c.
     \label{shear}
\end{multline}
   Now, combining (\ref{shear}) with (\ref{kurva1}) and plugging it into (\ref{kurva2}) the equation becomes
   \begin{multline}
       \dot{K}=-N\Big( \frac{1}{3}K^2 + 2\sigma^2 + 4 \pi G(\rho_{eff} + \mbox{tr} T^{eff})\Big) + D^aD_aN+N^aD_a K,
       \label{exkurvadot}
   \end{multline}
   where $\sigma^2 \equiv \sigma_{ab} \sigma^{ab}/2$.
Without losing generality, for convenience we will chose $N=1$ and $N^a=0$, then the metric (\ref{metrika}) becomes 
\begin{equation}
    ds^2=-dt^2 + h_{ab}dx^adx^b.
\end{equation}
The proper 3-dimensional volume in the hypersurface $\Sigma_t$ is simply $\det {h_{ab}}\equiv h$, so we can define the local scale factor as $a(t,x)^6=h$, which is the relative size of space measured by the Eulerian observer at each point, and can be thought as a generalisation of the usual scale factor in a FLRW spacetime. By contracting (\ref{exkurva}) with $h_{ab}$ one gets
\begin{equation}
h^{ab}K_{ab}=K=\frac{1}{\sqrt{h}}\frac{d}{Ndt}\sqrt{h},
\end{equation}
where we can introduce $d\tau = Ndt$, which is the proper time of the Eulerian observer definer by $x=const$. Then it follows
\begin{equation}
    K=\frac{3}{a}\frac{da}{d\tau}=\frac{3\dot{a}}{a} \qquad (\mbox{with }N=1) ,
\end{equation}
the field equation (\ref{exkurvadot}) becomes
\begin{equation}
    \ddot{a} + \Omega^2 a=0,
    \label{konacna}
\end{equation}
where:
\begin{equation}
    \Omega^2=\frac{1}{3}\Big( 2\sigma^2 + 4\pi G(\rho_{eff} + 3 p_{eff})\Big),
    \label{omega}
\end{equation}
where it was used $\mbox{tr} T^{eff}=h^{ab}T^{eff}_{ab}=3p_{eff}$. Finally, putting in equation (\ref{omega}) the modified terms $\rho_{mod}$ and $p_{mod}$ the equation takes the form
\begin{equation}
     \Omega^2=\frac{1}{3}\Big( 2\sigma^2 + 4\pi G(\rho_{mat}+\rho_{rad} + 3p_{rad}+\rho_{vac} + 3 p_{vac}) - \frac{\dot{S}}{2H}-S(t) \Big).
    \label{omega2}
\end{equation}

\section{The cosmological constant limit in the case of spacetime fluctuations}
It can be seen that by setting $S(t)=\lambda_{B}$  in equation (\ref{omega2}) the cosmological constant limit is restored. This limit was extensively studied in \cite{unruch},  where it was also taken that $\rho_{mat}=0$, $\rho_{rad}=0$ and $p_{rad}$. In this case it follows that $\lambda_{eff}=\lambda_{B}+ 8 \pi G \rho_{vac}<0$. This can be seen as follows. Taking the mentioned limit in equations (\ref{hamconstr}), (\ref{effrho}), (\ref{rhomod}) and considering an arbitrary Cauchy surface, it follows that the average spatial curvature is given by
\begin{equation}
\langle R^{(3)} \rangle_{\Sigma}=\lambda_{eff}+\langle K_{ab}K^{ab}- K^2 \rangle_{\Sigma}. 
\label{zakrivljenost}
\end{equation}
To be in accord with observations the average value of the spatial curvature needs to be very small. However, as discussed in \cite{unruch}, it can be shown that $\langle K_{ab}K^{ab}- K^2 \rangle_{\Sigma}$ leads to a very large positive contribution. It therefore follows that $\Lambda_{eff}$ needs to take large negative values in order to compensate this contribution. By inspecting the equation (\ref{omega2}), while taking into account the conclusion that $\lambda_{eff}=\lambda_{B}+ 8 \pi G \rho_{vac}<0$, we see that $\Omega^{2}$ must always be positive. Therefore, it follows from equation (\ref{konacna})
that $a$ needs to be an oscillatory function which crosses zero, as it basically represents the solution for the oscillator with varying frequency. In order to obtain the evolution equation for the shear, $\sigma$, appearing in (\ref{konacna}) one can use equation (\ref{shear}). Neglecting the average value of the spatial curvature, the average evolution of shear can be approximated with \cite{unruch}: 
\begin{equation}
\langle \sigma^{2}(\mathbf{x},t) \rangle=\sigma^{2}(\mathbf{x},0)e^{-2 \int K(\mathbf{x},t)'dt'},
\label{shear2}
\end{equation}
with the local expansion rate $K=\dot{a}/a$. It follows that the solutions for $a$ at the given spatial point, $\mathbf{x}$, represent oscillations with growing amplitude around the equilibrium point $a=0$ in the cosmological constant limit $S(t)=\lambda_{B}$. Since different points in space will oscillate with different phases, some regions of space will expand ($K>0$) and others will at the same time contract ($K<0$). Thus, when a macroscopic average is performed this will lead to a significant cancellation between the expanding and contracting contributions. Because of this, the effects of the original large value of cosmological constant can become hidden at macroscopic scales, leading to a small net expansion in accord with observations. \\ \\
The properties of solutions in the limit $S(t)=\lambda_{B}$ have, however, 
some unsatisfactory features. The first, milder, peculiarity is that the oscillations around zero imply the negative values of the local scale factor, $a$. Although it can be noted that the character of other physical quantities, like the spatial metric components, stays positive, one can remark that in cosmology $a$ plays a quite well defined physical role of the scale factor, defining a spatial physical distance between points - which would then become negative. \\ \\
The more serious limitation of the discussed solutions in the limit $S(t)=\lambda_{B}$ is that they lead to the physical singularity at the equilibrium point $a=0$. This of course directly contradicts the assumption 1. of our approach. The presence of singularities in cosmological solutions is, of course, not a new feature of this model, but can rather be viewed as a remnant of the classical general relativity, associated with the assumption $S(t)=const.$ However, the problem here gets even worse than in the classical FRWL model, since instead of one initial singularity, every point in space cyclically collapses and emerges from a singularity associated with the equilibrium point. For this reason we feel that the cosmological constant limit in this type of treatment of the space-time fluctuations is physically problematic. In the following sections we will show that going beyond this limit opens a much richer space of cosmological solutions: it is not only possible to have singularity-free oscillations with growing amplitude, but also solutions leading to a cosmological bounce and showing other types of cosmological evolution.

\section{The absence of oscillatory solutions around the cosmological bounce}
\begin{figure}[H]
    \centering
    \includegraphics[scale=0.75]{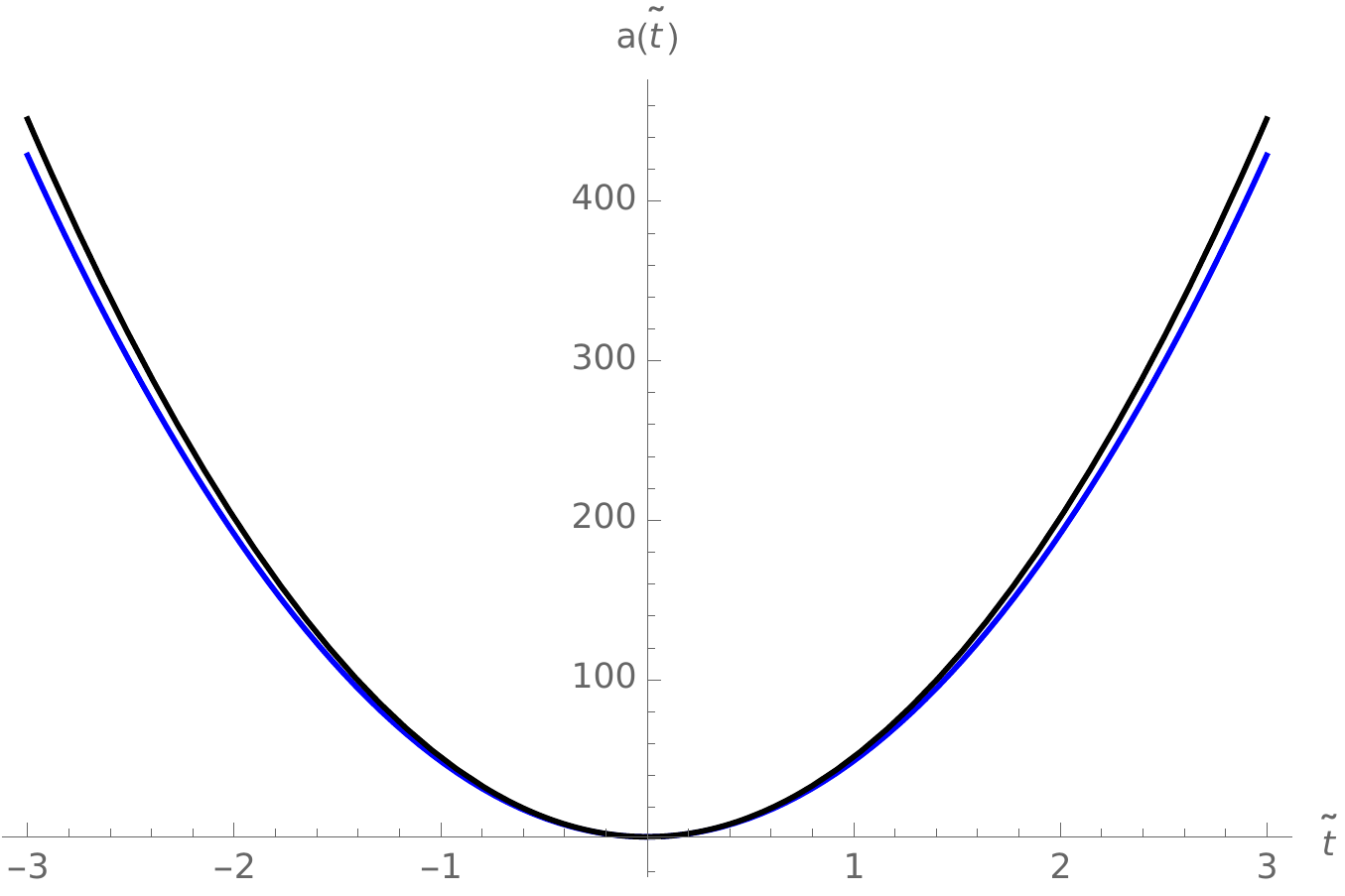}
    \caption{Here the scale factor obtained from the numerical solution (\ref{konacna}) is depicted in black while taking $S(t)$ given by (\ref{saroundbo}), while $a_{back}(t)=a_{min}+ C(t-t_b)^2/2$ is depicted in blue. We take $a_{min}=a(0)=1$, $\dot{a}(0)=0$ and $\rho^0_{mat}=\rho^0_{rad}=\rho_{vac}=1$. }
    \label{plot1}
\end{figure}

When the function $S(t)$ is no longer a constant, one cannot use equation (\ref{zakrivljenost}), with $\lambda_{eff}=S(t) + 8 \pi G \rho_{vac}$, to constrain its sign and magnitude for arbitrary periods in the future and past of the Universe based on observations -- since the average spatial curvature could in the very late or in the very early Universe in principle become arbitrary high. Therefore, the function $\lambda_{eff}(t)$ can in principle have arbitrary values and become both positive and negative during the evolution of the Universe, while approaching the large negative value today. We however note that the large negative value of $\lambda_{eff}$ is in accord with the requirements for the existence of a cosmological bounce at the level of classical (non-oscillating) FRWL spacetime, as we discussed in section 3. This could be a hint that the value of $S(t)$ is indeed also large and negative in the very early Universe. Furthermore, when one goes beyond the cosmological constant limit it is no longer true in general that $\Omega^ {2}>0$, since now the derivative of $S(t)$ appears in equation (\ref{omega2}) and therefore the function $\Omega^{2}$ can have different sings at different times. This opens the possibility for a much richer set of cosmological solutions than in the cosmological constant limit. \\ \\
We now want to study the cosmological bounce when the fluctuations of the space-time geometry are taken into account. 
The central assumption of this section will be that the scaling of the effective cosmological term with time, introduced by the corrections to the standard general relativity and discussed within the classical FRWL spacetime, will be, at least approximately, preserved when the quantum fluctuations of space-time are introduced. Following this assumption, our current task is to determine the consequences of solving equation (\ref{konacna}) instead of (\ref{modfri1}), while taking the time dependence of $S(t)$ to be given by equation (\ref{saroundbo}). We first proceed with the analytical discussion. With this aim, we use a perturbative picture, where we take that the background is given by the bounce spacetime in the absence of spacetime fluctuations and add the fluctuation effects at the next order. Thus at a single point $\mathbf{x}=\mathbf{x}'$ the background scale factor is given by $a_{back}(\mathbf{x'},t)=a_{min}+C(t-t_b)^2/2$. Using equation (\ref{omega2}) in this regime it can simply be shown that to the leading term around the bounce $\Omega^2$ is given by
\begin{equation}
    \Omega^2=-\frac{3C}{a_{min}}-\frac{8 \pi \rho^0_{rad}}{3a_{min}^4}+2\sigma^2+\mathcal{O}(t^2).
\end{equation}
Therefore, at a single spatial point, in order to have an absence of oscillatory solutions around the bounce we need to have $\Omega^2<0$ which yields
\begin{equation}
    \sigma^2<\frac{3C}{2a_{min}}+\frac{4 \pi \rho^0_{rad}}{3a_{min}^4}.
    \label{noscilation}
\end{equation}
We now confirm this conclusion by presenting the  numerical solution of equation (\ref{konacna}) in Fig. \ref{plot1} while taking $S(t)$ to be given by  (\ref{saroundbo}), with respect to the rescaled time parameter $\tilde{t}=(t-t_b)/t_b$.

\section{Singularity free cosmological solutions}
We now present the concrete solutions which do not suffer from pathological singularities. The central point of this section is to analyze solutions of equation (\ref{konacna}) by finding physically acceptable solution in the sense that the scale factor does not suffer from singularities $a(t_{singularity})=0$ and other pathological behaviour. Firstly, we begin with oscillatory solutions as a special case and a direct improvement of the main ideas presented in \cite{unruch}. The property of this solutions is that the metric indeed oscillates with growing amplitude, but without the scale factor touching zero and becoming negative. This will be more obvious in the subsec. 7.4, where it will be shown that the mathematical form of eq. (\ref{konacna}) does contain a non-homogeneous part, by the virtue of time-dependence in $S(t)$, which was not the case in \cite{unruch} and due to this, the points $a=0$ cannot be avoided in the cosmological constant limit.

\subsection{Examples of singularity free oscillations}

\subsubsection{Example of calculating the local scale factor, $a(t)$ from $S(a)$}

\begin{figure}[H]
    \centering
    \includegraphics[scale=0.29]{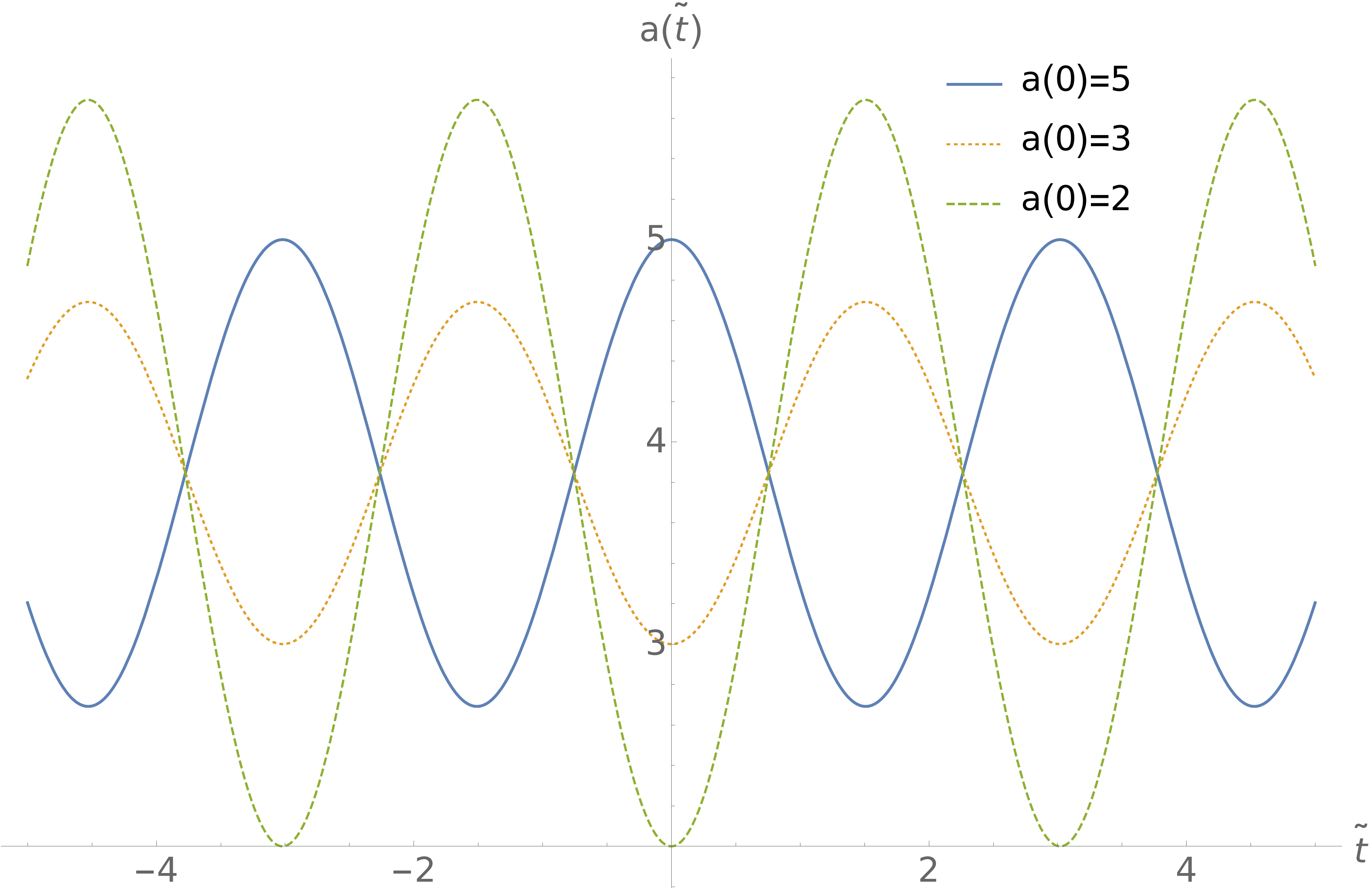}
    \caption{The scale factor is presented in this plot, with the parameters given by $\tilde{\lambda}_{c}=\tilde{K}+8\pi G\tilde{\rho}_{vac} =-1$, $\tilde{\sigma}_0=1$, $\tilde{c}=100$, with the different initial values: $a(0)=5$ depicted by the blue line, $a(0)=3$ orange line and $a(0)=2$ green line. The second Cauchy initial condition is $\dot{a}(0)=0$ for all configurations.}
    \label{oscilacije1}
\end{figure}

\begin{figure}[H]
    \centering
    \includegraphics[scale=0.32]{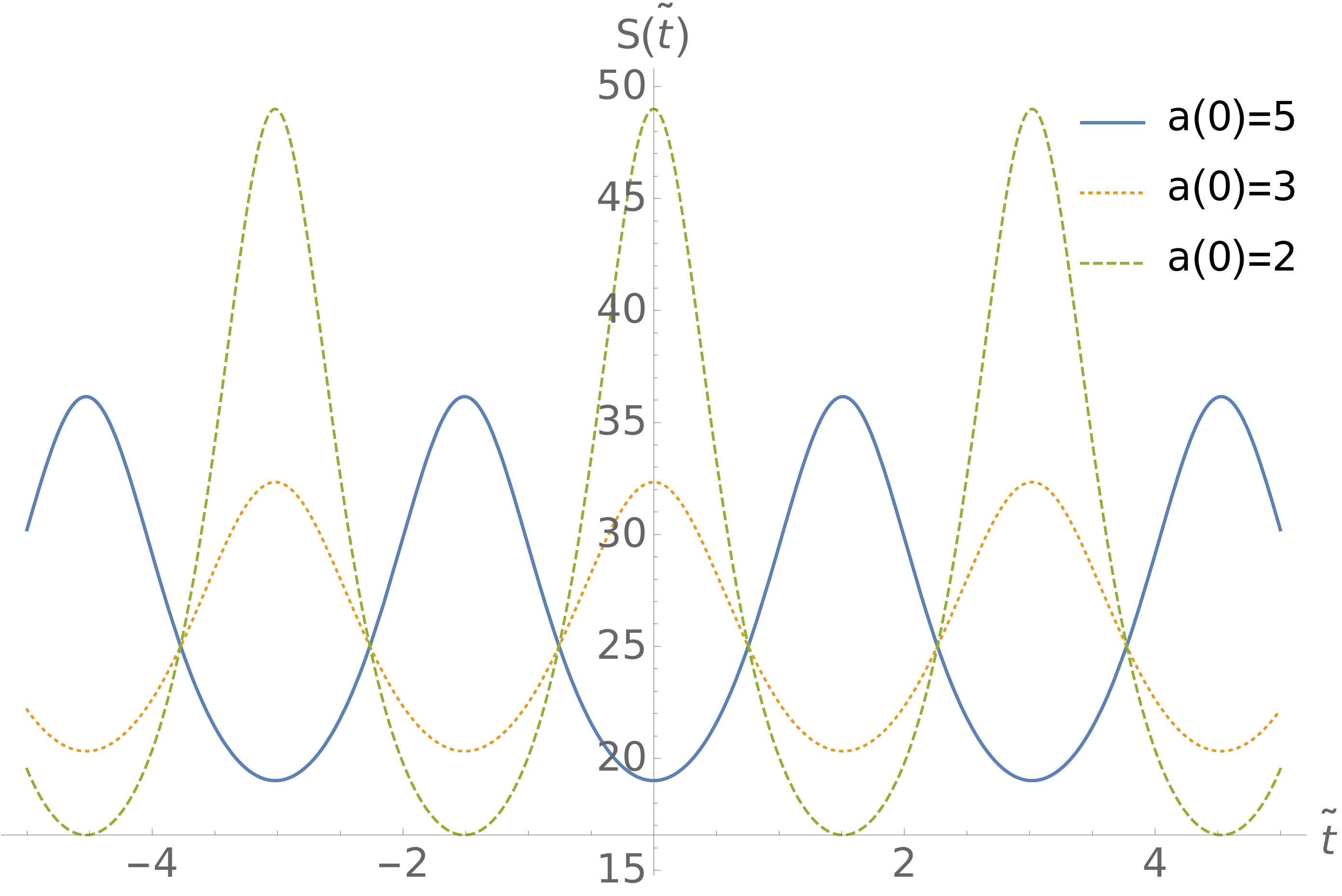}
    \caption{Here the function $S(t)$, calculated from the numerical solution of the scale factor where again the parameters are $\tilde{\lambda}_{c}=K+8\pi G\tilde{\rho}_{vac} =-1$, $\tilde{\sigma}_0=1$, $\tilde{c}=100$ with the different initial values: $a(0)=5$, is depicted by the blue line, $a(0)=3$ orange line and $a(0)=2$ green line. The second Cauchy initial condition is $\dot{a}(0)=0$ for all configurations.}
    \label{oscilacije2}
\end{figure}

\begin{figure}[H]
    \centering
    \includegraphics[scale=0.27]{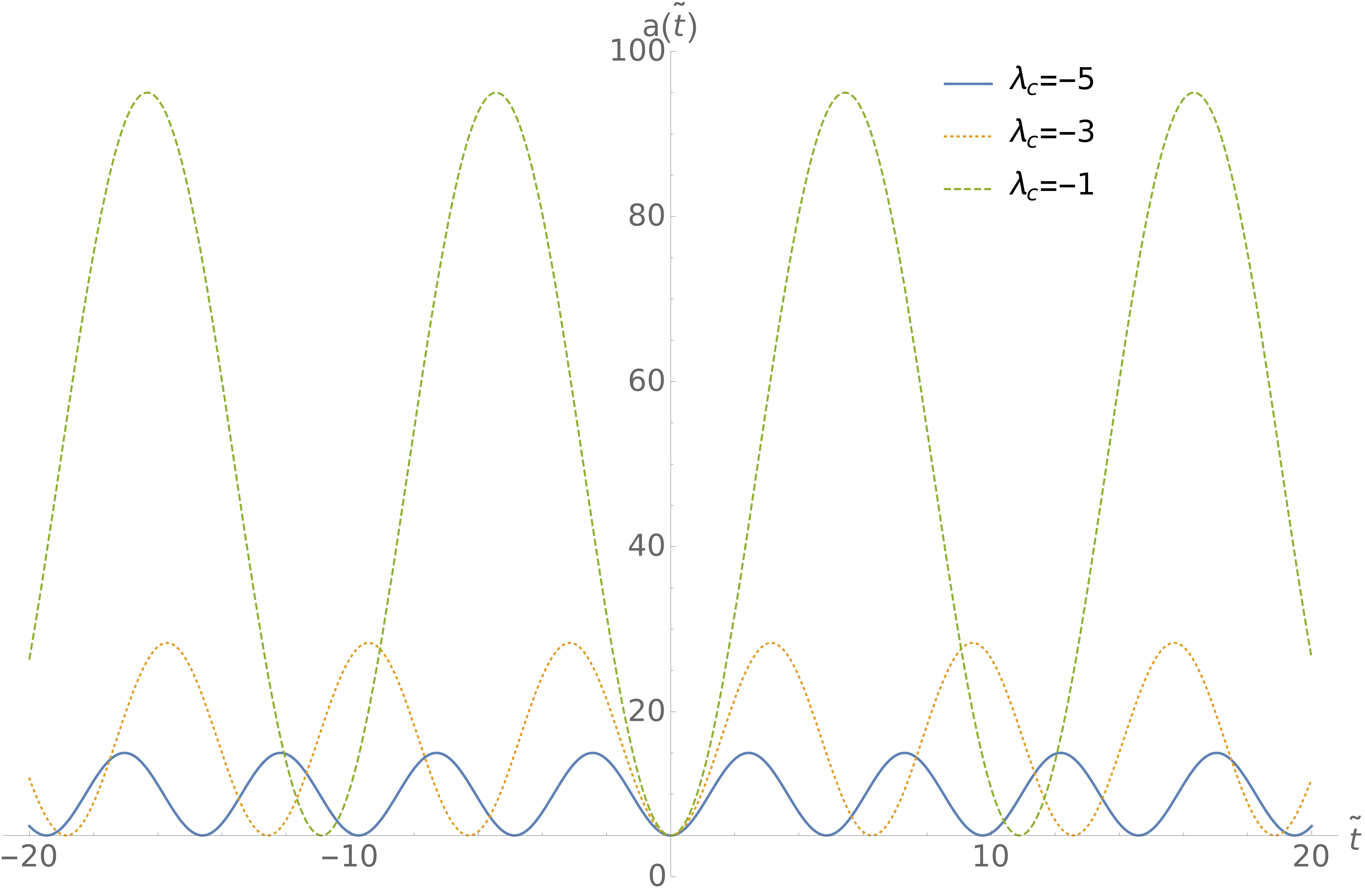}
    \caption{Here the scale factor $a(t)$ is depicted for different $\tilde{\lambda}_c$ values, the other constants are $\tilde{\sigma}_0=1$, $\tilde{c}=100$. The second Cauchy initial conditions are $a(0)=5$ and $\dot{a}(0)=0$ for all configurations.}
    \label{oscilacije3}
\end{figure}

In this subsection we present the singularity free oscillations by constructing the appropriate $S(t)$ function. An elegant way to enable the non-singular oscillations is to find the appropriate $S(a)$ in order to produce the constant non-homogeneity in eq. (\ref{konacna}). For simplicity we will exclude the matter and radiation contributions, in order to more clearly inspect the consequences achieved by introducing $S(t)$. As mentioned in Sec 5. the equation (\ref{konacna}) can be rewritten in a form similar to \cite{unruch}:
\begin{equation}
     \ddot{a} + \frac{1}{3}\Big( 2\sigma^2(a) -8\pi G \rho_{vac} -S(t) - \frac{\dot{S}}{2H} \Big) a=0.
     \label{konacna2}
\end{equation}
Let us, as before, consider this equation around a fixed spatial point
$\mathbf{x}=\mathbf{x'}$, where again 
from (\ref{shear2}) it follows
\begin{equation}
    \sigma(a)=\frac{\sigma_0}{a^3}.
\end{equation}
It can be seen from inspecting the above equation that in the case where the product $S(t)a(t)$ results in $S(t)a(t) \sim const.+ \omega^2 a(t)$ the equation (\ref{konacna}) has a very simple non-homogeneous part. Thus, in the simplest form this motivates the following choice
\begin{equation}
    S(a)=K+\frac{c}{a(t)},
\end{equation}
with constants $K$ and $c$.
Here, we also use the approximation for the shear evolution (\ref{shear2}) which we discussed earlier. We rescale the parameters to be more appropriate for numerical treatment:
\begin{equation}
    \tilde{t}=tH_0, \quad \tilde{\sigma}_0=\sigma_0 H_0^{-2}, \quad \tilde{\rho}_{vac}=\rho_{vac}H_0^{-2}, \quad \tilde{K}=KH_0^{-2}, \quad \tilde{c}=cH_0^{-2},
    \label{rescala}
\end{equation}
where $H_0$ is the Hubble parameter measured today.
By plugging this specific $S(t)$ in eq. (\ref{konacna}) with (\ref{omega2}) we can calculate the numerical solution which is depicted in Fig.\ref{oscilacije1} for different parameter values. Note that by changing the value of the constants involved: $\tilde{\lambda}_{c}=\tilde{K}+8\pi G\tilde{\rho}_{vac}$, not only the frequency is changed but also the amplitude, which comes from the nonlinear feature of equation (\ref{konacna}). The plot illustrating this is given in Fig. \ref{oscilacije3}. for different values of the  $\tilde{\lambda}_{c}$ constant.
It is also of interest to see the behaviour of $S(t)$ from the obtained solutions of $a(t)$ which is given in Fig.\ref{oscilacije2}

\subsubsection{Example of reconstruction of $S(t)$ from the local scale factor $a(t)$}

\begin{figure}[H]
\centering
  \includegraphics[width=0.48\linewidth]{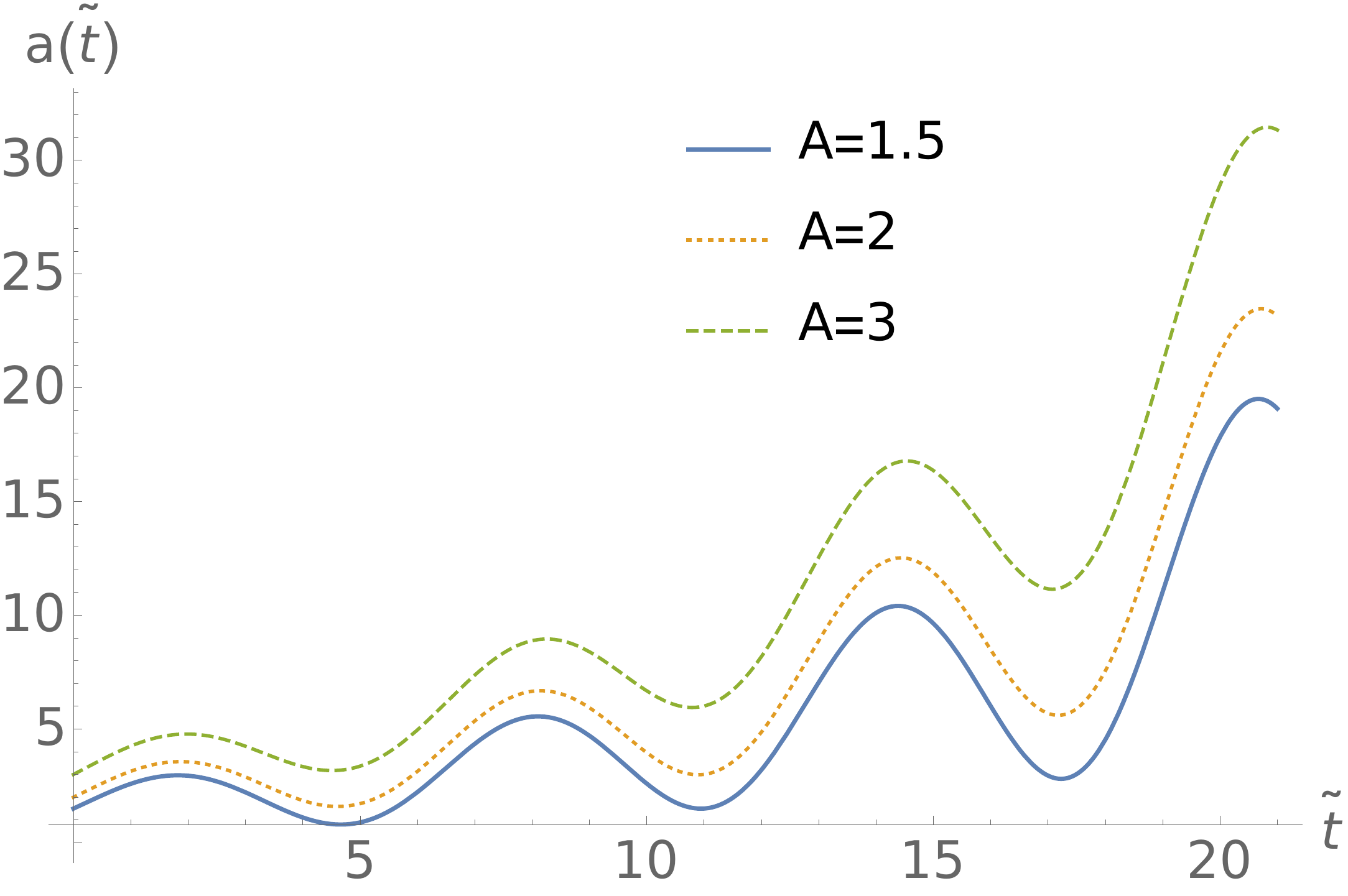}
  \includegraphics[width=0.48\linewidth]{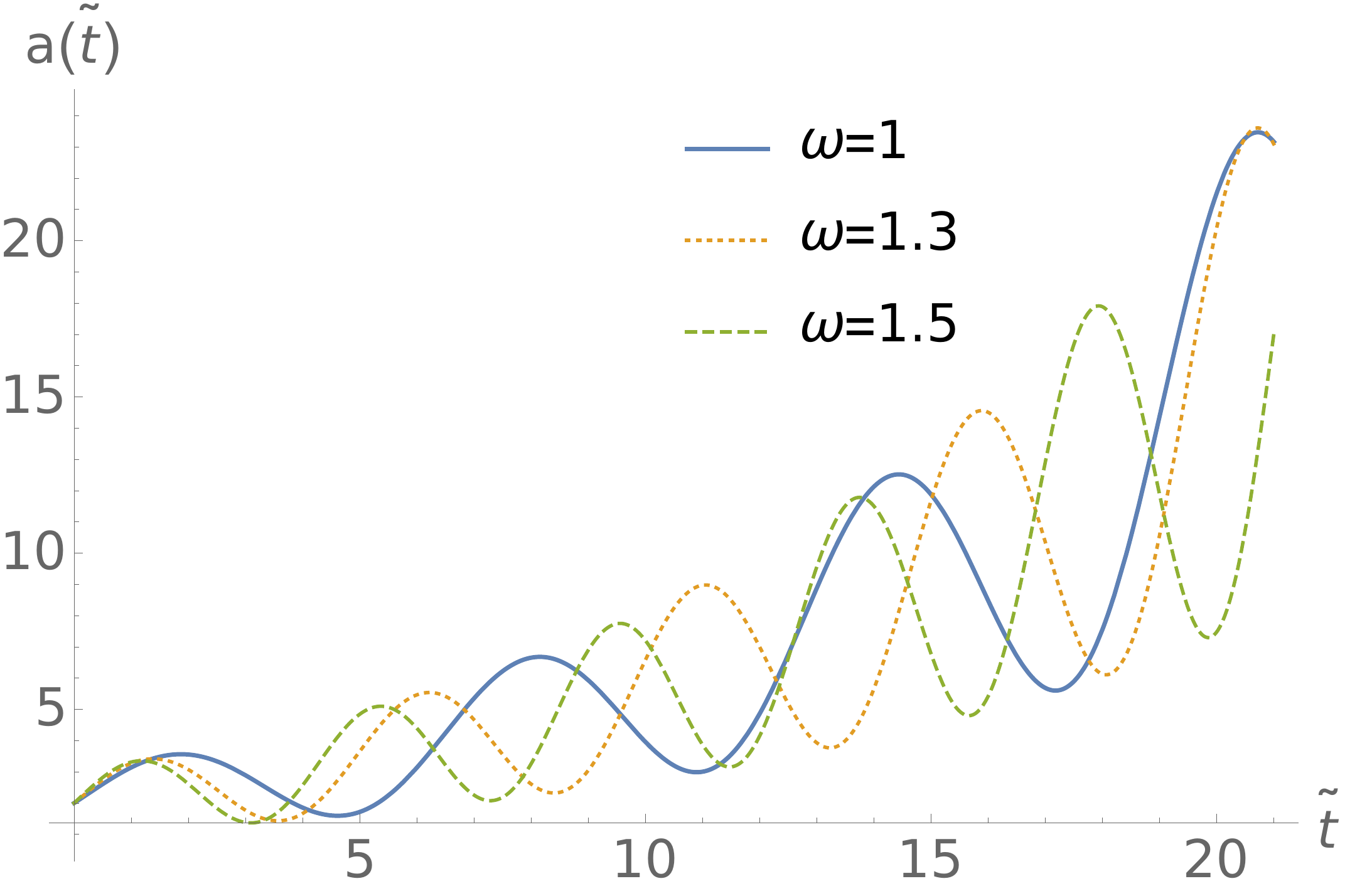}
  \caption{a) Example of the scale factor with rising amplitude with non singular oscillatory behaviour given by eq. (\ref{skalanum}). The parameters in this cases are $\tilde{\alpha}=0.1$ and $\tilde{\omega}=1$ with different parameter $A$.
  b) Example of the scale factor with rising amplitude with non singular oscillatory behaviour given by eq. \ref{skalanum}. The parameters in this cases are $\tilde{\alpha}=0.1$ and $A=2$ with different  frequency $\tilde{\omega}$.}
  \label{skalaprimjer}
  \end{figure}

\begin{figure}[H]
  \includegraphics[width=0.48\linewidth]{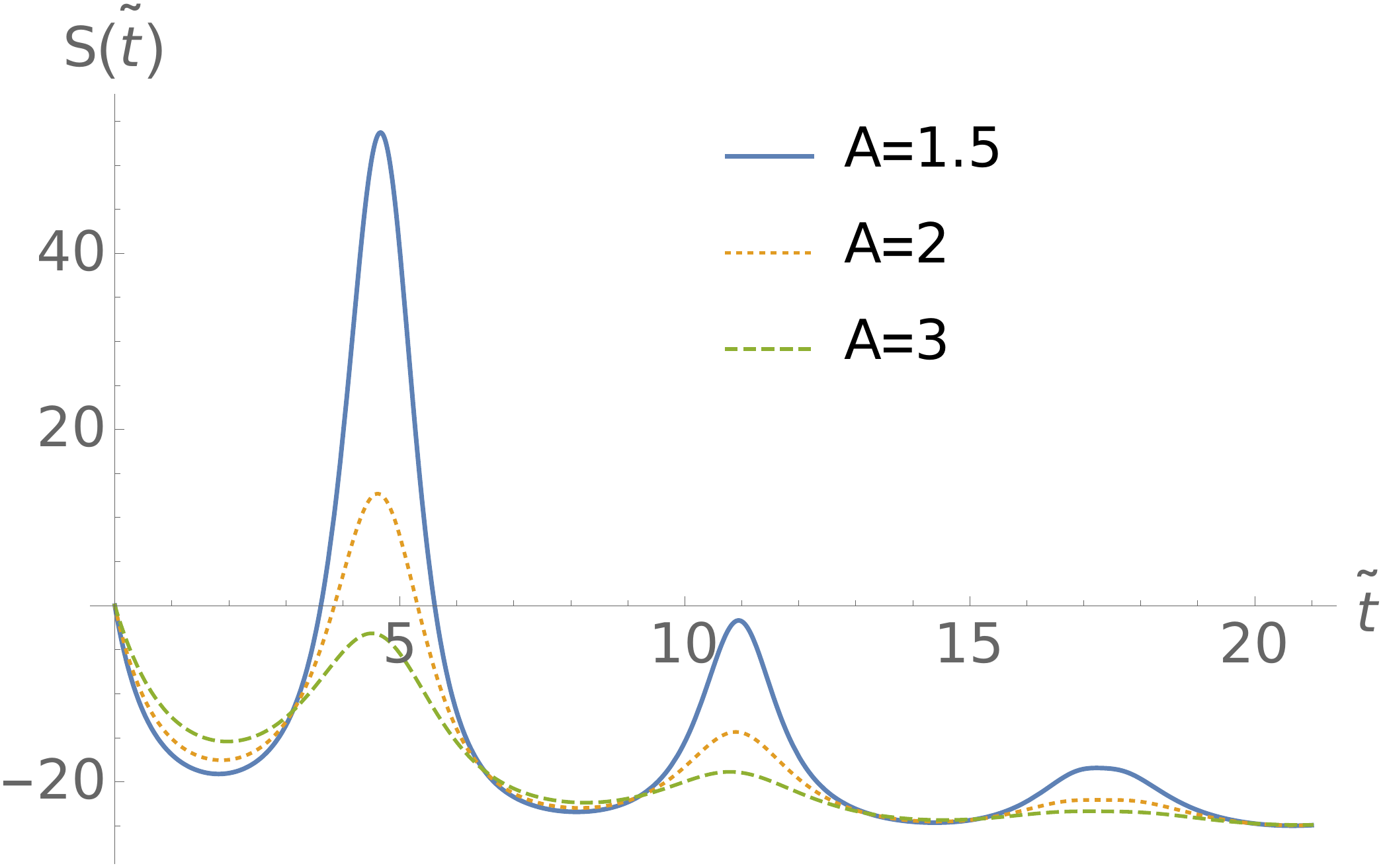}
  \includegraphics[width=0.48\linewidth]{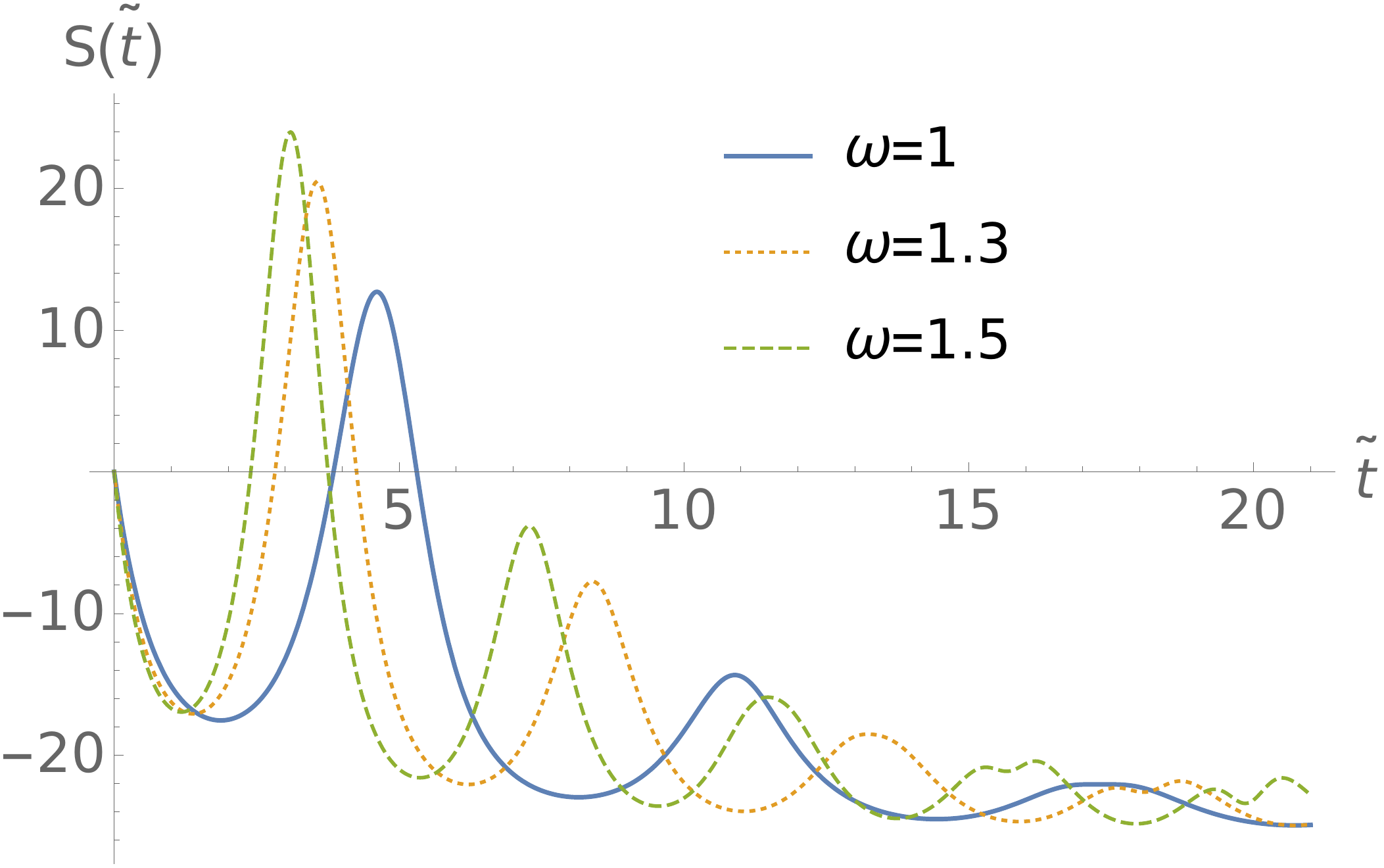}
  \caption{a) Numerical solutions for $S(t)$ given by oscillatory non singular scale factor from eq. (\ref{skalanum}). The parameters in this cases are $\tilde{\alpha}=0.1$ and $\tilde{\omega}=1$ with different parameter $A$.
   b) Numerical solutions for $S(t)$ given by oscillatory non singular scale factor from eq. \ref{skalanum}. The parameters in this cases are $\tilde{\alpha}=0.1$ and $A=2$ with different frequency $\tilde{\omega}$.}
  \label{sprimjer}
\end{figure}

Another way to construct specific $S(t)$ is to give $a(t)$ with the desired properties and then solve $S(t)$ from eq. (\ref{konacna2}) which is of the first order and linear in $S(t)$. As an example we choose:
\begin{equation}
    a(t)=Ae^{\alpha t}\Big(B+C \sin(\omega t)\Big).
    \label{skalanum}
\end{equation}
For simplicity we take $B=1/A^2$ and $C=1/A$ to inspect the three free parameters $\alpha$, $A$ and $\omega$.
Some of the desired test functions are given by eq. (\ref{skalanum}) and depicted in Fig. \ref{skalaprimjer}.
Plugging it into eq. (\ref{konacna2}) we can numerically extract the solutions of $S(t)$. Again, we rescale the parameters similarly as in the previous case:
\begin{equation}
      \tilde{t}=tH_0, \quad \tilde{\sigma}_0=\sigma_0 H_0^{-2}, \quad \tilde{\rho}_{vac}=\rho_{vac}H_0^{-2}, \quad \tilde{\alpha}=\alpha H_0^{-1}, \quad \tilde{\omega}=\omega H_0^{-1}
\end{equation}
In this specific cases the initial condition $S(0)=0$ is used. The numerical solutions are shown in Figs. \ref{sprimjer}.

\subsection{Other singularity free solutions}

\begin{figure}[H]
    \centering
    \includegraphics[scale=0.25]{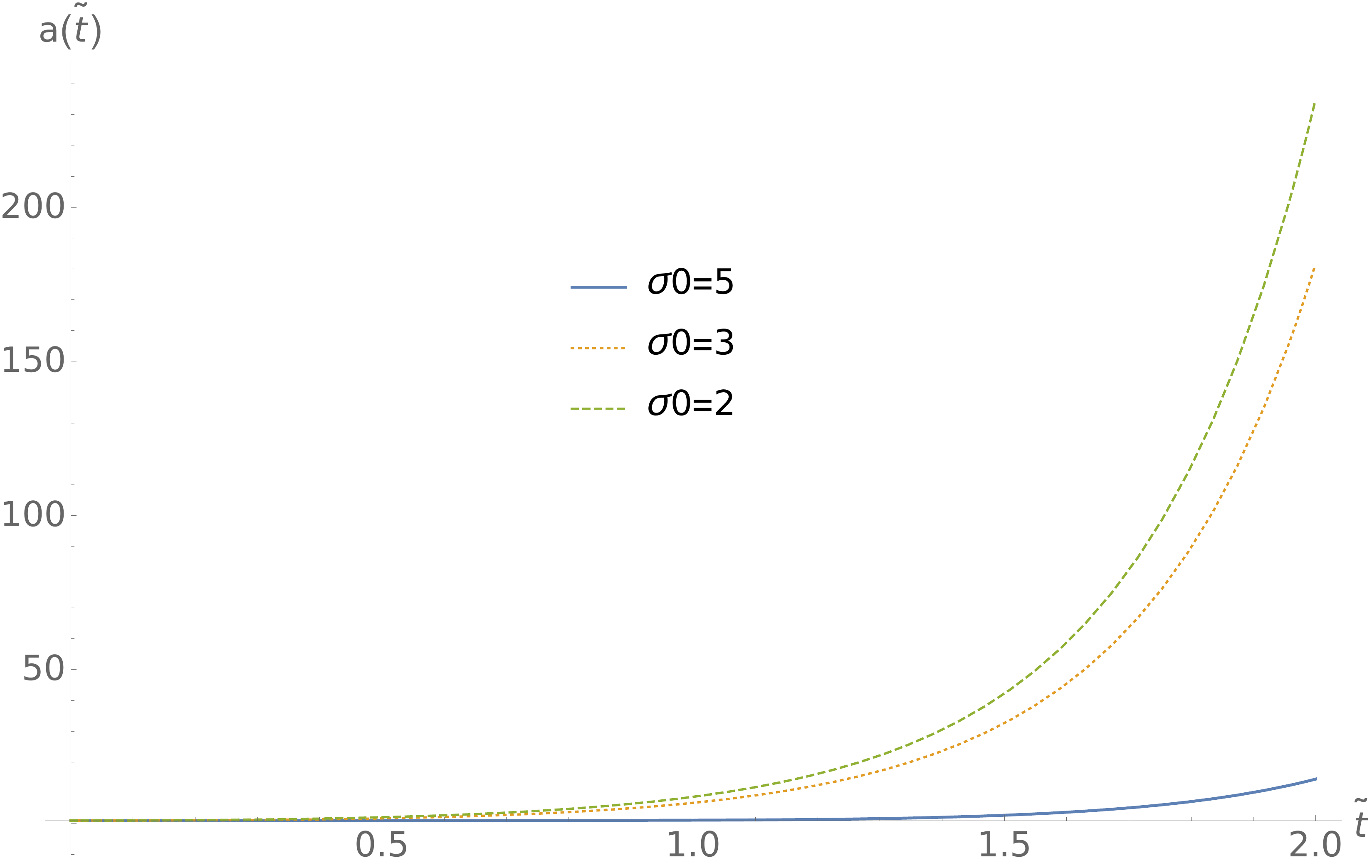}
    \caption{For a choice of the powers series function $S(t)=\tilde{A}\tilde{t}^3 +\tilde{B}\tilde{t}^2 + \tilde{C}\tilde{t}$  we calculate the scale factor $a(\tilde{t})$ for different parameters $\tilde{\sigma}_0$. The Cauchy initial condition are $a(0)=1$ and $\dot{a}(0)=0$ for all configurations.}
    \label{fignumskala1}
\end{figure}

In the previous section the oscillatory behaviour was preferred as a case of non-singular solutions which could be understood a direct improvement of \cite{unruch}. However, there is no mathematical necessity which would force such form of the local scale factor in the vicinity of the bounce, although  in the present era the scale factor should exponentially grow. Therefore, other non-singular solutions can be physically interesting and viable in this regime. Therefore, in this section we only investigate the possibility of non- pathological singular-free solutions in modified gravity and how to expect their modification in the effective contribution given by $S(t)$. 
Let us use a concrete example of the modifying term as a series
\begin{equation}
    S(t)=At^3+Bt^2+Ct,
    \label{modterm1}
\end{equation}
where $A$, $B$ and $C$ are positive constants. Again, using eq. (\ref{konacna2}):
\begin{equation}
     \ddot{a} + \frac{1}{3}\Big( 2\sigma^2 -8\pi G \rho_{vac} -S(t) - \frac{\dot{S}}{2H} \Big) a=0,
\end{equation}
we can numerically calculate the local scale factor. The rescaling in this case is given similarly as in (\ref{rescala}) with the addition
\begin{equation}
     \tilde{A}=A H_0^{-3}, \qquad \tilde{B}=B H_0^{-2}, \qquad \tilde{C}=CH_0^{-1}
\end{equation}
For simplicity we use $\tilde{A}=\tilde{B}=\tilde{C}=\tilde{\rho}_{vac}=1$ and neglect matter and radiation energy density including only vacuum energy density. The solutions are given in fig. \ref{fignumskala1}. with different values of the shear factor $\tilde{\sigma}_0$.
For the power series choice (\ref{modterm1}) the solutions have a greater growth by decreasing $\tilde{\sigma_0}$. This could potentially be due to the oscillatory behaviour of the shear component in the differential equations. Furthermore, by decreasing $\sigma_0$ the shear is suppressed and only the  effective vacuum energy density, namely $S(t)$, gives the main contribution which increases with time.

\subsection{Loop quantum gravity motivated solutions}
\begin{figure}[H]
    \centering
    \includegraphics[scale=0.25]{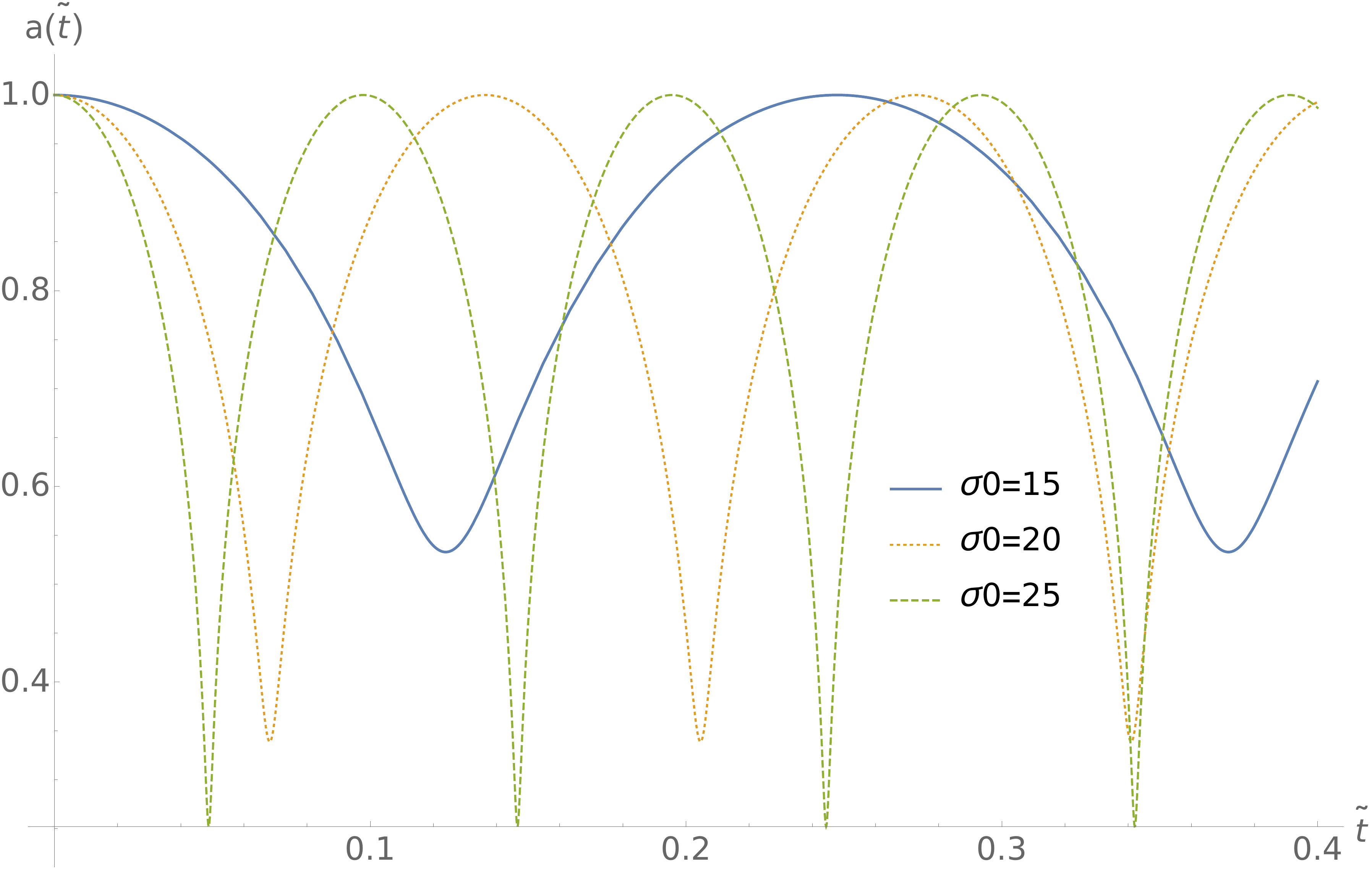}
    \caption{For LQC the function $S(t)$ becomes $S(t)=-8\pi G \rho^2/\rho_c$, from which we calculate the scale factor $a(\tilde{t})$ for different parameters $\sigma_0$. The  Cauchy initial condition are $a(0)=1$ and $\dot{a}(0)=0$ for all configurations.}
    \label{lqca}
\end{figure}

\begin{figure}[H]
    \centering
    \includegraphics[scale=0.25]{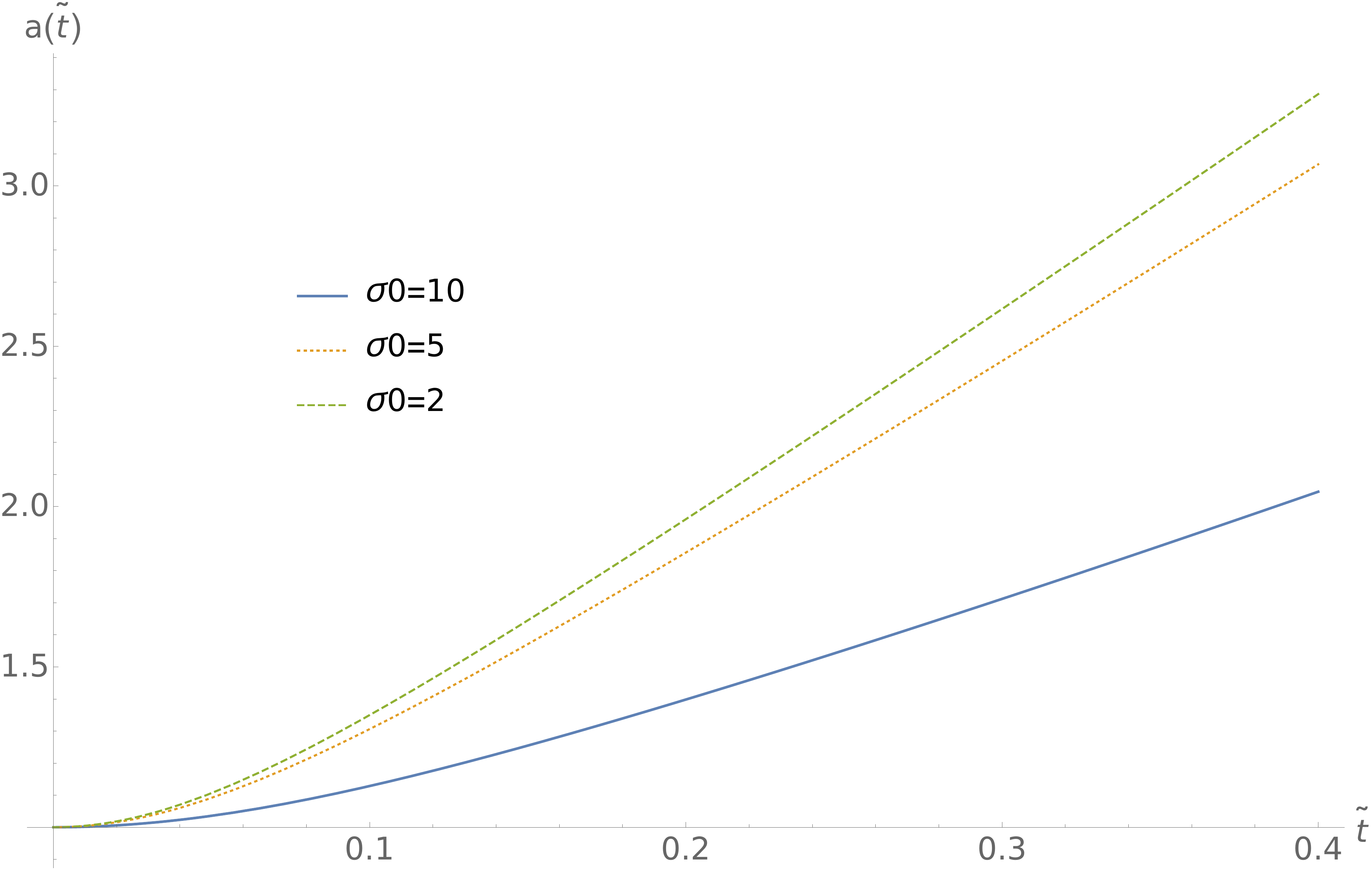}
    \caption{Here the scale factor $a(\tilde{t})$ is computed for $\tilde{\sigma}_0<10$ where the oscillatory behaviour cease to exist. The  Cauchy initial condition are $a(0)=1$ and $\dot{a}(0)=0$ for all configurations.}
    \label{lqca2}
\end{figure}

In the quest of quantizing gravitational interaction there are several approaches. Two of the most prominent are canonical and covariant approaches. From the canonical approach \cite{dewitt, mtw, thiemann1}, among other proposals, a theory called Loop quantum gravity (LQG) arises \cite{rovelli1, rovelli2, nicolai, nicolai2, astekar, astekar2, astekar3}, while the most well known program arising out of the covariant approach is perhaps the group of theories sometimes referred as M-theories and string theories.  \cite{struna1, witten}. The whole set of this theories can be applied to the cosmological setting and in principle the effective contributions of this set can be expressed by the term $S(t)$. However, from this whole set of quantum gravity theories there is a sub-theory specifically established in the context of cosmology which arises from LQG and is known as Loop quantum cosmology (LQC) \cite{agullo, miranda, martin}. It should be remarked that LQC is quite successful in describing the inflatory problem in a natural way from the principles of LQG applied in the early universe. In this sense it is interesting to explore the behaviour of equations of motion described by the effective modification arisen from LQC. By a simple transformation 
\begin{equation}
    S(t)=-8 \pi G\frac{\rho^2}{\rho_c},
\end{equation}
the effective equation for LQC is obtained, where $\rho$ is the matter energy density and $\rho_c$ is the critical parameter of the theory \cite{miranda, singht}. The equation of interest can be rewritten
\begin{equation}
    \ddot{a}+\frac{1}{3}\Bigg(2 \sigma(a)^2 + 4 \pi (\rho + 3p) + \frac{8 \pi}{ \rho_c}\Big( \frac{4 \dot{a}}{a}\rho\frac{d \rho}{dt} + \rho^2\Big) \Bigg)=0.
    \label{lqceq1}
\end{equation}
Taking into account  matter, radiation and $\rho_{vac}$, and assuming their classical evolution (thus ignoring their fluctuations), $\rho$ can be expressed as a function of $a$
\begin{equation}
\rho(a)=\frac{\rho^0_{rad}}{a^4}+\frac{\rho^0_{mat}}{a^3} + \rho_{vac},
\end{equation}
where (\ref{lqceq1}) became the second order differential equation for $a$ which we solve numerically. Firstly, we re scale  variables and parameters as 
\begin{equation}
    a(\tilde{t})=a(t), \qquad \tilde{t}=t \rho_c^{1/2}, \qquad \tilde{\rho}=\rho/\rho_c, \qquad \tilde{\sigma}=\sigma \rho_c^{-1/2}.
\end{equation}
The initial conditions are specified as $a(0)=1$ and $\dot{a}(0)=0$, the other parameters for simplicity are $\rho_{vac}/\rho_c=1$, $\rho^0_{mat}/\rho_c=1$ and $\rho^0_{rad}/\rho_c=1$.
The solution is plotted in fig \ref{lqca} for different parameters $\tilde{\sigma}_0$.
It can be seen that the effective frequency is increased by increasing $\tilde{\sigma}_0$, also we found a critical $\tilde{\sigma}_0 \sim 10$ where the oscillatory solutions cease to exist. Above this limit the solution of $a(t)$ is monotonically increasing as can be seen in fig \ref{lqca2}.
We conclude that including the contribution of LQC in the resulting effective action can lead to the non-singular behaviour of the scale factor. Depending on the shear parameter $\sigma_0$ different behaviours can be obtained: the oscillatory solution is obtained from $\sigma_0 \gtrsim10$ (similar to the Unruch solution) and monotonically increasing function of the local scale factor results for $\sigma_0   \lesssim 10$. In this formalism there is a numerical evidence that LQC is consistent and compatible with cosmological solutions without singularities in the given regime.

\subsection{Analytical reconstruction of singularity free solutions}
Finally, it is possible to reconstruct analytical singularity free solutions from the main equation of motion (\ref{konacna}). This equation can be rewritten in the form
\begin{equation}
   \ddot{a} \pm \omega^2 a=f(t), 
   \label{skaladif}
\end{equation}
where
\begin{equation}
    \omega^2=\frac{8\pi G}{3}\rho_{vac},
\end{equation}
$\pm$ sign depends on the chosen sign of $\rho_{vac}$. It is obvious that eq. (\ref{skaladif}) is essentially a nonlinear second order differential equation in $a(t)$, however for the purpose of analytically proving  the non-singular behaviour of $a(t)$, taking all the combination on the right hand side to be just a function of time, one can rewrite the equation as a second order linear equation with the non-homogeneous part given by
\begin{equation}
    f(t)=-\frac{2}{3}\sigma(a)^2a+\frac{1}{6} \frac{\dot{S} a^2}{\dot{a}} + \frac{S}{3}a-\frac{4\pi G}{3}a(\rho_{mat} + \rho_{rad}+3p_{rad}).
    \label{tjeranje}
\end{equation}
Now, formally the general solution of $a(t)$ is given by
\begin{equation}
    a(t)=A \sin(\omega t) + B \cos (\omega t) + a_{p_{+}} \qquad \rho_{vac}>0,
\end{equation}
\begin{equation}
     a(t)=A \sinh(\omega t) + B \cosh (\omega t)+ a_{p_{-}} \qquad \rho_{vac}<0,
\end{equation}
with respectively
\begin{equation}
    a_{p_{+}}=\frac{1}{\omega} \int_0^t f(t') \sin\Big(\omega(t-t')\Big)dt'
\end{equation}
\begin{equation}
    a_{p_{-}}=\frac{1}{\omega} \int_0^t f(t') \sinh\Big(\omega(t-t')\Big)dt',
\end{equation}
where $A$ and $B$ are integration constants.
The general solution can be given only by specifying the $f(t)$ function in order to calculate the particular solution. But, some general conclusions on the scale factor can be given by this procedure. The equation \ref{tjeranje} can be rewritten in the form
\begin{equation}
    \dot{S}+\frac{2\dot{a}}{a}S=6\frac{\dot{a}}{a^2}f(t)+
    4\sigma^2\frac{\dot{a}}{a}+8\pi G(\rho_{mat}+2\rho_{rad})\frac{\dot{a}}{a},
\end{equation}
then by using
\begin{equation}
    \sigma(a)=\frac{\sigma_0}{a^3}, \qquad \rho_{mat}(a)=\frac{\rho_{mat}^0}{a^3}, \qquad \rho_{rad}(a)=\frac{\rho_{rad}^0}{a^4},
\end{equation}
the general solution of $S(t)$ can be written as
\begin{equation}
    S(t)=\frac{C}{a^2}- \frac{\sigma_0^2}{a^6} - 8\pi G \Big( \frac{\rho_{mat}^0}{a^3} + \frac{\rho_{rad}^2}{a^4}\Big) +\frac{1}{a^2}\int6 f(t) \dot{a}dt.
\end{equation}
It is now clear that the non-homogeneous part of (\ref{skaladif}) is necessary in order to have singularity free solutions, otherwise the scale factor will always cross the zero point. Also, from the form of equation  (\ref{skaladif}) one can simply analytically construct singularity-free consistent solutions. As an example lets use a simple case
\begin{equation}
    \rho_{vac}>0, \qquad f(t)=f_0, \qquad a(0)=a_{min}, \qquad \dot{a}(0)=0, \qquad \frac{f_0}{\omega^2}>a_{min},
\end{equation}
the last condition keeps the scale factor positive for all $t$. The general solution of the scale factor is then
\begin{equation}
    a(t)=\Big( a_{min}-\frac{f_0}{\omega^2}\Big) \cos(\omega t ) + \frac{f_0}{\omega^2},
\end{equation}
the consistent solution of $S(t)$ is 
\begin{equation}
    S(t)=\frac{C}{a^2}- \frac{\sigma_0^2}{a^6} - 8\pi G \Big( \frac{\rho_{mat}^0}{a^3} + \frac{\rho_{rad}^2}{a^4}\Big) +\frac{6f_0}{a^2}\Big( a_{min} - \frac{f_0}{\omega^2} \cos(\omega t) \Big).
\end{equation}
On the other hand with $\rho_{vac}<0$
\begin{equation}
     \rho_{vac}<0, \qquad f(t)=f_0, \qquad a(0)=a_{min}, \qquad \dot{a}(0)=0, \qquad \frac{f_0}{\omega^2}<a_{min},
\end{equation}
now the last inequality is changed in order to keep the scale factor positive without crossing zero point, then the given solutions are
\begin{equation}
    a(t)=\Big( a_{min}-\frac{f_0}{\omega^2}\Big) \cosh(\omega t ) + \frac{f_0}{\omega^2},
\end{equation}
\begin{equation}
    S(t)=\frac{C}{a^2}- \frac{\sigma_0^2}{a^6} - 8\pi G \Big( \frac{\rho_{mat}^0}{a^3} + \frac{\rho_{rad}^0}{a^4}\Big) +\frac{6f_0}{a^2}\Big( a_{min} - \frac{f_0}{\omega^2} \cosh(\omega t) \Big),
\end{equation}
which is indeed no-nsingular for all $t$. It is important to note that $f(t)$ cannot be zero in which case the particular solution is zero, consequently the scale factor will cross zero point. Interestingly, the modifying function $S(t)$ is proportional to $1/a^2$, therefore as the scale factor approaches zero the modifying terms become more important. This is expected as the effective quantum contributions are the leading contributions in this regimes.

\section{Discussion and conclusion}
The extensive work on modified gravity theories and exotic forms of stress-energy tensor in the past decades was motivated mostly by the dark energy problem -- but also other problems of standard general relativity, such as the existence of singularities. As the works conducted within this research program explored the possibilities and observational consequences of models beyond general relativity they, at the same time, investigated the potential routs to quantum gravity. The usual approach in the exploration of physics beyond general relativity consists of choosing a specific type of modification of action integral for gravity (or some type of exotic stress-energy source) and then elaborating its consequences. The crucial point of this approach is ,,the lucky guess" of the appropriate Lagrangian, which can lead to desired type of solutions. The main difficulty in research on modified gravity at this moment is the absence of direct and relevant observational clues which would clearly point to the physics beyond general relativity. In this situation, thousands of different modified models were developed, accompanied with vast research literature focused on the elaboration of their physical consequences. The main problem is that the properties of such solutions are strongly dependent on series of technical assumptions, which are to a large extent physically non-motivated and can not be empirically verified at the moment. It is thus not clear which properties of solutions are more general and which are only specific for a concrete model. When it comes to the fundamental questions -- such as: do singularities really exist in Nature -- it is not even possible to sketch a clear approach towards their answer, since it is not at all clear which of the thousands of potential models of modified gravity, if any, is actually preferred by Nature. This signals the need for developing alternative approaches for the research in modified gravity, not in order to abandon the standard one, but in order to approach the problem from different angles.  
\\ \\ 
In this work we have, following some ideas presented in \cite{mi1}, proposed a complementary approach. Instead of choosing a specific type of modification of the gravitational Lagrangian, we explored a very general type of modification constrained by the condition that it leads to the non-singular evolution of the Universe. This general type of Lagrangian modification leads to the effective corrections which enter field equations for gravity, $G_{\mu \nu}^{eff}$, restricted only by the stress-energy conservation. Then, instead of focusing on the general space-time, we consider the application to the space-time of interest, in this case the FRWL space-time. The high degree of symmetries of this space-time leads to a very simple form of the modification term, since it needs to be a function only of the time coordinate, and can thus be written as a single function of one variable, $S(t)$. Postulating the necessary properties of cosmological solutions  -- such as the absence of cosmological singularities and the consistency with the general features of observations, we discuss the basic properties of the modification with respect to the standard general relativity, modelled by the function $S(t)$. The disadvantage of this approach is that the application of this method is limited to only the specific space-time geometry under consideration, as is the the FRWL spacetime considered in this paper. However, the clear advantages of this approach is that the equations and their solutions are model independent and much more general,
including most of the known types of modified gravity theories. 
\\ \\
Taking this approach, we have first simply demonstrated that the cosmological constant fine tuning problem is absent under quite a broad set of conditions. Namely, if the function $S(t)$, representing the corrections to the classical Einstein's equation of FRWL space-time, is a continuous function of time, leads to the cosmological bounce, and at some late time approaches some value $S_{asymp}> 3H_{0}^{2} - 8 \pi G \rho_{vac}$ (with $H_{0}$ being the value of the Hubble parameter today)) then the corresponding cosmological solutions will include  the phase in which the measured effective cosmological term $\lambda_{effective}= S(t) + 8 \pi G \rho_{vac}$ reaches the arbitrary small positive values. In this context, what appears as the mysterious fine tuning of the two unrelated constants in the standard cosmological model, comes as a natural consequence of the fact that $S(t)$ needs to support the cosmological bounce and smoothly evolve to its late-time value. The large negative values taken by $S(t)$ are in accord with both the requirement for the cosmological bounce and the compensation of the large value of $\rho_{vac}$. Without further determination of the details of the possible $S(t)$ evolution, which requires further theoretical and observational investigation, it is not possible to give technical details of this picture, such as the duration of the interval of matter and radiation equality or get a feeling of "how small" the cosmological term, $\lambda_{eff}(t_{0})$, actually is at this time, since for this it needs to be compared with its maximum and minimum value during the cosmological evolution. \\ \\
Here, an objection could be made: "in order to obtain the small cosmological constant which is observed, in the framework presented here the term $S(t)$ needs to be large at late times. Therefore, the problem of the cosmological constant is just shifted to the problem of large value of this term at late times". However, this objection is obviously not justified. The large negative value of $S(t)$ is not chosen artificially, but it is shown to be necessary for supporting the bounce. Combining this condition for the existence of the cosmological bounce and assumption that $S(t)$ continues to grow in some arbitrary fashion at late times after the current cosmological moment (while we showed that this term needed to grow between the bounce and the late time acceleration phase), we have proven that there will necessary exist a moment where $\lambda_{effective}$ will reach arbitrary small values. Thus, in this framework there is no mysterious fine tuning of two unrelated constants on numerous decimal places, but just a compensation between a function and a constant around a specific point of time, following from the presented discussion. The question "why would $S(t)$ continue to rise at late times" is certainly not a part of the fine tuning problem of the cosmological constant and this situation just represents one possible type of the late-time cosmological evolution in the modified gravity theories. This question is therefore out of the scope of this work. We can, however, note that such type of the late cosmological evolution could be required in cyclical cosmological models, in order to solve the problems of stability during the contraction phase \cite{ mi2}. 
\\ \\
We then considered the effect of space-time fluctuations on the previously discussed setting. To do this, we started from the very general 
metric \ref{metrika}, modelling the fluctuations of space-time,  which cause it to lose the high degrees of symmetries present in the FRWL spacetime. After performing the $3+1$ decomposition of this space-time, by considering the modified Einstein's equations we finally arrive at the field equation (\ref{konacna}) which resembles the corresponding field equation presented in \cite{unruch}, but now also containing the function $S(t)$ and its derivative, modelling the departure from the standard Einstein's field equations. The simplifying assumption taken was that the spatial variations of the modifying  function $S$ can be ignored, so that this function can be treated as dependent only on the time coordinate. We first show that, if the initial value of the shear stays small enough to be in accordance with \ref{noscilation}, it will not lead to qualitative changes around the bounce. Furthermore, due to modification of the field equations for gravity, the necessary properties of solutions change drastically: instead of the necessity of oscillating solutions which need to cross zero and thus to have singularity at $a=0$ in the $S(t)=const.$ limit, we show that it is now possible to have various types of non-singular solutions, including the ones which are oscillatory and non-singular. Such solutions were constructed and discussed in section 7. We also considered the solutions motivated by the LQC correction and found that they are consistent with the absence of singularities and having a proper evolution for the cosmological constant problem resolution. At the end of this section a discussion on the analytical reconstruction of singularity free solutions is also provided. We note that these solutions, discussed in section 7, and the ones around the bounce are corresponding to the very different regimes of the very early and the late Universe. Those regimes can then of course be matched by an appropriate interpolation functions, with the constraint that they should lead to all of the confirmed $\Lambda$CDM properties. Such discussion is beyond the scope of this work, and it should be based on the observational data from which the satisfactory fits on $S(t)$ should be made.   \\ \\
We have thus followed the approach in which what is called "the observed cosmological constant" is actually understood to consist of two contributions: i) the vacuum energy density and ii) the effects of departure from the standard field equations from gravity (and is thus in general time dependent). As we have discussed, assuming the FRWL spacetime, this can be written as: $\lambda_{eff}=S(t) + 8 \pi G \rho_{vac}$. We have first shown that a large negative value reached by $S(t)$ in the early Universe can support the cosmological bounce, and thus assuming the existence of the bounce, naturally explains the currently observed small value of $\lambda_{eff}$, under the assumption that $S(t)$ evolves towards much less negative (or potentially even positive at some period in the future) values in the late Universe. When the FRWL spacetime is replaced by the much more general metric, in order to describe quantum fluctuations of the spacetime itself, then invoking the modifications to Einstein's equation can also further hide the huge value of the vacuum energy in the spacetime fluctuations, while fixing the singularities otherwise appearing in the solutions. \\ \\
Further research work within this approach should also be focused on observationally reconstructing and constraining the possible forms of $S(t)$ function from the cosmological data, needed for developing further conclusions on the possible past and future evolution of our Universe.


\begin{thebibliography}{}
\bibitem{tests}
Relativistic Cosmology (Cambridge University Press, Cambridge, England,)
G.F.R. Ellis, R. Maartens, M.A.H. MacCallum
\bibitem{hawk1}
S. Hawking, Proc.Roy.Soc.Lond.A 294 (1966) 511-521 
\bibitem{hawk2}
S. Hawking, Proc.Roy.Soc.Lond.A 295 (1966) 490-493
\bibitem{hawk3}
S. Hawking, Proc.Roy.Soc.Lond.A 300 (1967) 187-201
\bibitem{dm1}
K. A. Olive et al. (Particle Data Group), Chin. Phys. C, 38 090001 (2014)
31
\bibitem{dm2}
Planck Collaboration, P. A. R. Ade, N. Aghanim, C. Armitage-Caplan, et al., Astron.Astrophys 571
A16 (2014)
\bibitem{linde}
A. Linde, Lect.Notes Phys. 738: 1-54 (2008)
\bibitem{k1}
 Steven  Weinberg.   The  cosmological  constant  problem.Rev. Mod. Phys., 61:1–23, Jan 1989.
\bibitem{k2}
 A.  D.  Dolgov.The  Problem  of  vacuum  energy  andcosmology.  In Phase  transitions  in  cosmology.  Proceedings, 4th Cosmology Colloquium, Euroconference, Paris,France, June 4-9, 1997, 1997.
 \bibitem{k3}
Weinberg,  “Theories  of  the  cosmological  constant,”  in:   N. Turok, Critical  Dialogues  in  Cosmology(World Scientific, Singapore, 1997), p. 195, arXiv:astro-ph/9610044.
\bibitem{k4}
A. Padilla, arXiv:1502.05296 [hep-th]
\bibitem{stefancic}
H, Stefancic, Phys.Lett.B 670 (2009) 246-253
\bibitem{sola}
J. Solà, A. Gómez-Valent, Int.J.Mod.Phys.D 24 (2015) 1541003
\bibitem{sola2}
 J. Sola, J. Phys. Conf. Ser. 453 (2013) 012015
 \bibitem{carlip}
S. Carlip, Phys. Rev. Lett. 123, 13, 131302 (2019)
 \bibitem{unruch}
Q. Wang, W. G. Unruh,
Physical Review D 102, 023537 (2020)
 \bibitem{unruch2}
  Q. Wang, Z. Zhu, W. G. Unruh,
    Physical Review D 95, 103504 (2017)
  \bibitem{unruch3}
 S. S. Cree, T.  M. Davis, T. C. Ralph, Q.  Wang, Z. Zhu, W. G. Unruh,
Physical Review D  98, 063506 (2018)
\bibitem{wheeler}
 John A. Wheeler.  On the Nature of quantum geometro-dynamics.Annals Phys., 2:604–614, 1957.
 \bibitem{hidekazu}
 Hidekazu Nariai,
 Progress of Theoretical Physics, Vol. 46, No. 2, 1971
 \bibitem{ratbay}
 R. Myrzakulov, L. Sebastiani, S. Zerbini, 
 International Journal of Modern Physics D, v.22, N.8, 1330017 (2013)
 \bibitem{bojowald}
 M. Bojowald, AIP Conf.Proc. 910 (2007) 1, 294-333
 \bibitem{nojiri}
 S. Nojiri, S.D. Odintsov, V.K. Oikonomou, 
Phys.Rept. 692 (2017) 1-104
\bibitem{mudrac}
A. DeBenedictis, D. Horvat, S. Ilijic, S. Kloster, K. S. Viswanathan, 
Class.Quant.Grav. 23 (2006) 2303-2316
\bibitem{cai}
 Y. F. Cai, S. H. Chen, J. B. Dent, S. Dutta and E. N. Saridakis, Class.Quantum Grav. 28, 215011(2011)
 \bibitem{roshan}
 M. Roshan, F. Shojai, Physical Review D 94, 044002 (2016)
 \bibitem{salehi}
 A. Salehi, M. Mahmoudi-Fard, Eur.Phys.J. C78 (2018) no.3, 232
 \bibitem{pan}
 S. Pan, Mod.Phys.Lett. A33 (2018) no.01, 1850003
 \bibitem{sahoo}
  P. Sahoo, S. Bhattacharjee, S.K. Tripathy, P.K. Sahoo, Mod.Phys.Lett.A35 (2020) 2050095
 \bibitem{bari}
 P. Bari, K. Bhattacharya, S. Chakraborty, Universe 4 (2018) no.10, 105
 \bibitem{bajardi}
 F. Bajardi, D. Vernieri, S. Capozziello, 
 Eur. Phys. J. Plus 135, no.11, 912 (2020)
 \bibitem{mi1}
 P.Pavlovic, M. Sossich,
Physical Review D 103, 023529 (2021)
\bibitem{mi2}
P. Pavlovic, M. Sossich,
Physical Review D  95, 103519 (2017)
\bibitem{clifton}
T. Clifton, P. G. Ferreira, A. Padilla, C. Skordis, 
Physics Reports 513, 1 (2012), 1-189
\bibitem{natty}
N. Leite, P. Pavlovic, 
Classical and Quantum Gravity 35, 21, 215005 (2018)
\bibitem{dewitt}
Bryce S. DeWitt. Quantum theory of gravity. i. the canonical theory. Physical
Review, 160(5):1113, (1967)
\bibitem{mtw}
R. Arnowitt, S. Deser, and C. W. Misner. Consistency of the canonical reduction
of general relativity. Journal of Mathematical Physics, 1(5):434–439, (1960)
\bibitem{thiemann1}
Thomas Thiemann. Modern canonical quantum general relativity. Cambridge
University Press, (2008)
\bibitem{rovelli1}
Carlo Rovelli and Francesca Vidotto. Covariant Loop Quantum Gravity. Cambridge Monographs on Mathematical Physics. Cambridge University Press, (2014)
\bibitem{rovelli2}
Carlo Rovelli. Quantum Gravity. Cambridge Monographs on Mathematical
Physics. Cambridge University Press, November (2004)
\bibitem{nicolai}
Hermann Nicolai, Kasper Peeters, and Marija Zamaklar. Loop quantum gravity:
an outside view. Classical and Quantum Gravity, 22(19):R193, (2005)
\bibitem{nicolai2}
Hermann Nicolai and Kasper Peeters. Loop and spin foam quantum gravity: A
brief guide for beginners. In Approaches to fundamental physics, pages 151–184.
Springer, (2007)
\bibitem{astekar}
Jorge Pullin and Abby Ashtekar, editors. Loop Quantum Gravity, the first 30
years, volume 4 of 100 years of general relativity. World Scientific, (2017)
\bibitem{astekar2}
Abhay Ashtekar and Jerzy Lewandowski. Background independent quantum
gravity: A Status report. Class. Quant. Grav., 21:R53, (2004)
\bibitem{astekar3}
Abhay Ashtekar, Eugenio Bianchi, Rep. Prog. Phys. 84, 042001 (2021)
\bibitem{struna1}
Becker, Katrin; Becker, Melanie; Schwarz, John  String theory and M-theory: A modern introduction. Cambridge University Press (2007)
\bibitem{witten}
Witten, Edward "Fivebranes and knots". Quantum Topology. 3 (1): 1–137 (2012)
\bibitem{agullo}
Ivan Agullo, Parampreet Singh, World Scientific series "100 Years of General Relativity."
\bibitem{miranda}
Marcello Miranda, Daniele Vernieri, Salvatore Capozziello, Francisco S. N. Lobo, Eur. Phys. J. C 81, 975 (2021)
\bibitem{martin}
Martin Bojowald, Living Reviews in Relativity volume 8, 11 (2005) 
\bibitem{singht}
Parampreet Singh, S. K. Soni, Class. Quant. Grav 33 (2016) 125001
\bibitem{struyve}
Ward Struyve, Loop quantum cosmology and singularities. Sci Rep 7, 8161 (2017)

\end{thebibliography}
\end{document}